\documentclass[preprint,12pt,times,authoryear]{article}

\usepackage{natbib}
\setcitestyle{open={(},close={)}}
\usepackage{amsmath}

\usepackage{amssymb}
\usepackage{multirow}
\usepackage{bm}
\usepackage{comment}
\usepackage{rotating}
\usepackage[english]{babel}
\usepackage{array}
\usepackage{lscape}

\usepackage{makeidx}
\usepackage[doublespacing]{setspace}
\usepackage{threeparttable}
\usepackage{longtable,booktabs}
\usepackage{float}

\usepackage{graphicx}
\usepackage{caption}
\usepackage{subcaption} % for subfigures

\usepackage{color}
\usepackage[usenames,dvipsnames]{xcolor}
\usepackage[]{hyperref}
\usepackage{url}

\definecolor{darkblue}{rgb}{0.0,0.0,0.6}

\hypersetup{colorlinks=true,pdfborder={0 0 0},linkcolor=black,citecolor=darkblue}

\usepackage{tgpagella}

\newdimen \dummy
\dummy=\oddsidemargin
\title{What does Network Analysis teach us about International Environmental Cooperation?\footnote{We thank Tony Bolsen, Robert Falkner, Brigitte Granville, Steffen Kalbekken, Hakon Saelen and Marco Vincenzi for their useful comments on a previous version of the paper. We are also grateful to Jim Alayo-Arnabat and Isabella Cingolani for excellent research assistance in assembling the dataset. Carattini, Fankhauser and Gennaioli acknowledge support from the Grantham Foundation for the Protection of the Environment and from the UK Economic and Social Research Council through the ESRC Centre for Climate Change Economics and Policy. Carattini further acknowledges support from the the Swiss National Science Foundation, grant number PZ00P1\_180006$/$1. Fankhauser further acknowledged support from the University of Oxford Strategic Research Fund. The usual disclaimer applies. Corresponding author: sam.fankhauser@ouce.ox.ac.uk}
\vspace{2mm}}% \\ \vspace{0.5cm}\small Preliminary draft}

\author{Stefano Carattini\footnote{Department of Economics, Andrew Young School of Policy Studies, Georgia State University}\;\footnote{CESifo}\;\footnote{Grantham Research Institute on Climate Change, London School of Economics}\;\footnote{Department of Economics and SIAW-HSG, University of St. Gallen}\;, Sam Fankhauser\footnotemark[4]\;\footnote{Smith School of Enterprise and the Environment, University of Oxford}\;, Jianjian Gao\footnote{School of Business and Management, Queen Mary University of London}\;\;, \\ Caterina Gennaioli\footnotemark[7]\;\;, and Pietro Panzarasa\footnotemark[7]}

\date{\vspace{3mm} \today \vspace{-5mm}}

\begin{document}

%\begin{frontmatter}

\maketitle

\newpage

\begin{abstract}
Over the past 70 years, the number of international environmental agreements (IEAs) has increased substantially, highlighting their prominent role in environmental governance. This paper applies the toolkit of network analysis to identify the  network properties of international environmental cooperation based on 546 IEAs signed between 1948 and 2015. We identify four stylised facts that offer topological corroboration for some key themes in the IEA literature. First, we find that a statistically significant cooperation network did not emerge until the early 1970, but since then the network has grown continuously in strength, resulting in higher connectivity and intensity of cooperation between signatory countries. Second, over time the network has become closer, denser and more cohesive, allowing more effective policy coordination and knowledge diffusion. Third, the network, while global, has a noticeable European imprint: initially the United Kingdom and more recently France and Germany have been the most strategic players to broker environmental cooperation. Fourth, international environmental coordination started with the management of fisheries and the sea, but is now most intense on waste and hazardous substances. The network of air and atmosphere treaties is weaker on a number of metrics and lacks the hierarchical structure found in other networks. It is the only network whose topological properties are shaped significantly by UN-sponsored treaties.

\end{abstract}
\vspace{0.5cm} 


%\vspace{0.7cm} \textbf{JEL}:
%\end{frontmatter}

%\setcounter{page}{1}

\newpage

\section{Introduction}
In a globalised and interconnected world, international cooperation is crucial to the betterment of society. Pressing environmental issues are a particular case in point. Many of the most urgent environmental dilemmas require collaboration across countries~ \citep{rosenau2004governance, andonova2009transnational, o2017environment}. Sometimes cooperation involves a relatively limited number of parties (e.g. to manage a shared water body), sometimes it requires broad coalitions of many nations (e.g. for global threats like climate change).

Understanding how environmental coalitions have emerged and expanded is therefore an important question in international cooperation and global governance research. The literature has tackled the problem both theoretically and empirically, using among others the tools of game theory \citep[e.g.][]{barrett2003environment, barrett2007cooperate, de2015international,10.1111/jeea.12138, doi:10.1086/684478, o2017environment}, international relations \citep[e.g.][]{falkner2013nation,carlsnaes2002handbook} and experimental economics \citep[e.g.][]{milinski2006stabilizing, milinski2008collective, tavoni2011inequality, barrett2012climate}. 

The subject of interest in these studies is typically a particular international environmental agreement (IEA). Researchers are interested in the political, game theoretic or behavioural dynamics that explain the emergence, design or effectiveness of a treaty \citep[e.g.][]{breitmeier2011effectiveness, young1999effectiveness}. The success of the Montreal Protocol on Substances that Deplete the Ozone Layer, for example, has been explained by the rapid emergence of a low-cost substitute to chlorofluorocarbons, which created an economic incentive to cooperate \citep{barrett1994self,wagner2016estimating}.

What tends to be overlooked by the focus on individual treaties is that, as a collective, IEAs have given rise to a dense network of environmental cooperation. Recent decades have witnessed a significant increase in the number of IEAs, reaching a total of almost $2000$ in 2015. The number of signatories has increased from $6$ in 1869 (when there were fewer sovereign nations) to $238$ in 2015, including not just nation states, but also international organisations, dependent territories and sub-national entities. As such, IEAs are a central building block in the global system of multilateral, multi-level, polycentric environmental governance~\citep{ostrom2009polycentric, jordan2018governing}.

The breadth and depth of environmental cooperation through IEAs has been documented in information sources such as ECOLEX \citep{ECOLEX} and the International Environmental Agreements Data Base \citep{mitchell2003international, mitchell2020we}. Their main interest is often the classification and categorisation of individual treaty types. 

Here, we are interested in the network of cooperation these treaties create. While every environmental agreement, or attempt to lead to one, has its own particular motivation and objectives, our ability to successfully manage global environmental threats depends on the synergies and interactions between multiple treaties. The strength of cooperative ties is affected not only by country attributes \citep{carlsnaes2002handbook}, but also by the structure of cooperation \citep{kinne2013network}. 

The structure of a network \textit{per se} provides important insights into its functioning as a system of interacting components ~\citep{jackson2010social}.  Many important mechanisms that determine the likelihood of cooperation, such as reciprocity and reputation \citep{dai2010international, hafner2009network}, are related to third-party ties. As such the network of cooperation not only reflects existing cooperative relationships, but also influences the costs and benefits of future cooperative attempts \citep{kinne2013network}.

We apply the toolkit of network analysis to ECOLEX, one of the largest collections of data on IEAs, with a view to better understand the structure and dynamics of global environmental cooperation. In doing so, we elucidate, with new quantitative evidence, some long-standing debates in the economics and political science literature on international (environmental) agreements, global governance, and international cooperation and offer topological corroboration for several conjectures supported by either qualitative or preliminary correlational evidence. We shed light on how collective and cooperative environmental behaviour emerged and evolved over time, and unveil the role that individual countries played in sustaining or hindering such collaborative behaviour.     

Specifically, we analyse the topology of an inter-temporal environmental cooperation network, where each node is a country that has signed IEAs and each link connects two countries that have co-signed one or more IEAs. Our data cover 546 environmental treaties agreed between 1948 and 2015. Crucially, the topological position of a country in the network is not simply based on the number of treaties it has signed, but is assessed by constructing a proper null model and filtering out connections that might also be found in a network of random connections.

Our research approach is inductive. Rather than trying to test particular hypotheses or governance theories, we let the data speak. We use global topological metrics to identify a number of salient network features - or stylised facts - that describe the evolution and current landscape of international environmental cooperation.

We identify four such stylised facts. First, network analysis dates the emergence of meaningful environmental cooperation to the early 1970s. A statistically significant environmental cooperation network materialised in 1971 and reached stability in 1980. Before then, treaty links were too weak. Since then the network has grown steadily in size and strength, resulting in higher connectivity between signatory countries. As such, our data support earlier findings on the pivotal role played by events like the 1972 UN Conference on the Human Environment in Stockholm, as posited in \citet{falkner2019emergence}.

Second, network analysis shows how, through membership interconnections, environmental cooperation has become closer, denser and more cohesive. The paths through which countries can reach each other have shortened, creating more effective platforms for policy coordination and knowledge diffusion. This is consistent with the view of IEAs as vehicles for engagement, which provide organisational structures, reflect a shared purpose and engender trust ~\citep[e.g.,][]{meyer1997structuring,bernauer2010comparison, ostrom2009polycentric, owen_trust_2008}.

Third, we find that the environmental cooperation network, while global, has a noticeable European imprint. Initially the United Kingdom and more recently France and Germany have been the most important network nodes, through which IEAs are facilitated. Although there are fluctuations, the strategic position of countries has remained relatively stable over time. As such, network analysis corroborates the view of international relations scholars like \citet {vogler2007european} and \citet{kelemen2010trading} who discuss the leadership role of European countries in (domestic and international) environmental issues, explaining it as a reflection of domestic political dynamics (e.g. the role of the Green Party in Germany) and a consequence of the intermittent interest of the United States in environmental diplomacy. 

Fourth, we find that international environmental coordination started with the management of fisheries and the sea, but is now most intense on waste and hazardous substances. The networks on species, waste and natural resources have a hierarchical structure, which is absent in the networks on sea and fisheries and air and atmosphere. Despite its policy salience, the network of air and atmosphere treaties is comparatively less cohesive and intense. It is also the only subject area where UN-sponsored treaties (such as those on climate change and trans-boundary air pollution in Europe) have had a significant impact on the topological properties of the network. The results speak to the level of 'regime complexity' described by \citet{meyer1997structuring} and \citet{keohane_regime_2011}, and might explain the ambivalence towards the UN in much of the environmental governance literature \citep{biermann2004assessing, ivanova2010unep, mee2005role}. 

Our findings are relevant and contribute to a number major debates in the economics and political science literature. At its broadest, our paper is therefore part of the wider theoretical and conceptual literature in economics and political science on (environmental) governance and international (environmental) cooperation.

From a methodological perspective the paper relates more narrowly to a strand of empirical literature at the crossroad of economics and political science, which leverages large data sets on IEAs, such as the one we use, to identify empirical patterns of environmental cooperation. Four studies, which our paper complements, are worth highlighting.

\cite{kim2013emergent} examines a network of IEAs linked through citations and finds an international environmental governance system that is characterised by a cohesive polycentric legal structure. \cite{hollway2016multilevel} apply network analysis to the governance of global fisheries, using and identifying a high degree of social embeddedness in the system. \cite{wagner2016estimating} uses a structural model of international negotiations to estimate the date when countries ratified the Montreal Protocol as well as the dynamics of trade agreements. Finally, ~\cite{mitchell2020we} discuss the potential, without yet exploiting it, of the International Environmental Agreements Data Base, a similar database to ECOLEX, to better understand the formation of IEAs. 

Our paper also relates to a connected literature, which applies network analysis to wider international relations contexts, including trade, financial integration, and technology diffusion \citep[e.g.][]{smith1992structure, kim2002longitudinal, fagiolo2010evolution, schiavo2010international, vega2018technology, htwe2020coevolution, hafner2009network}.

The remainder of the paper is organised as follows. Section \ref{Sec-methodology} describes the data and the construction of the environmental cooperation network. Section \ref {Sec:motivation} motivates the subsequent analysis with a set of simple statistics about international environmental cooperation. The main results are contained in sections \ref{sec:size_and_connectivity} to \ref{sec-results_by_treaties}, each of which introduces a different stylised fact, or salient network feature. Section \ref{Sec:Conclusions} concludes.

\section{Data and methodology}\label{Sec-methodology}

\subsection{Environmental treaty data}

We use global data on IEAs from ECOLEX \citep{ECOLEX}, which combines information on environmental laws and treaties from several sources. As in \cite{mitchell2003international}, the treaties included in the ECOLEX database are defined as \textit{intergovernmental documents intended as legally binding with a primary stated purpose of preventing or managing human impacts on natural resources}\footnote{The official definition for international treaties originates from the Vienna Convention on The Law of Treaties (1969). The definition used here has been adapted to treaties on environmental matters.}. 

The database contains information on $1,998$ environmental treaties signed by $238$ parties between 1868 to 2015. We exclude $1,411$ treaties on which important dates (e.g., on treaty ratification and entry into force) are missing. In addition, we focus on treaties signed by countries and not by other parties such as international organisations, dependent territories and sub-state territories. Finally, we focus on the post-war period. After this selection, our final sample comprises 546 environmental treaties signed over the period 1948-2015 by $200$ countries. The 546 treaties are illustrative of the network as a whole and include the largest and most important global treaties.

For each treaty we have information on signatory countries, subject areas (e.g., nature conservation, waste, climate, etc.), the date it was signed and the date it entered into force. The data also include country information on the dates of treaty ratification, acceptance or approval and the date of withdrawal, where applicable. 

The data are then organised into a country-treaty-year panel, which lists for each country the IEAs it was a member of at the end of each year and/or for each treaty its signatories at the end of each year. 

\subsection{Network construction}\label{subsection: method}

We next convert the country-treaty-year data into a sequence of annual environmental cooperation networks. The conversion process involves three steps. The first step is the construction of annual bipartite networks, that is, two-mode networks where a link is established between a country and a treaty if the former has signed the latter. In the second step, we use one-mode projections to convert the bipartite networks into one-mode cooperation networks, in which a link is established between any two countries if both have signed at least one common treaty. 

The final step concerns statistical validation, that is, the identification of statistically significant links through comparison with an appropriate null model.  

\subsubsection{The bipartite networks}

Our raw data document which country is a member of which treaty in a specific year. In network analysis, this type of data is called affiliation data. A broad range of affiliation data has been studied, such as women's attendance of events~\citep{davis1941deep}, corporate board memberships~\citep{robins2004small, battiston2004statistical}, co-authorship data~\citep{newman2001scientifica, newman2001scientificb}, and actors-movies relations~\citep{watts1998collective, newman2001scientifica}. Co-membership of groups or events, such as countries' membership in IEAs, indicates social ties or in our case cooperative relationships among countries signing the same treaties ~\citep{borgatti2011analyzing}.

\begin{figure}[H]
 \centering
 \includegraphics[scale=0.45]{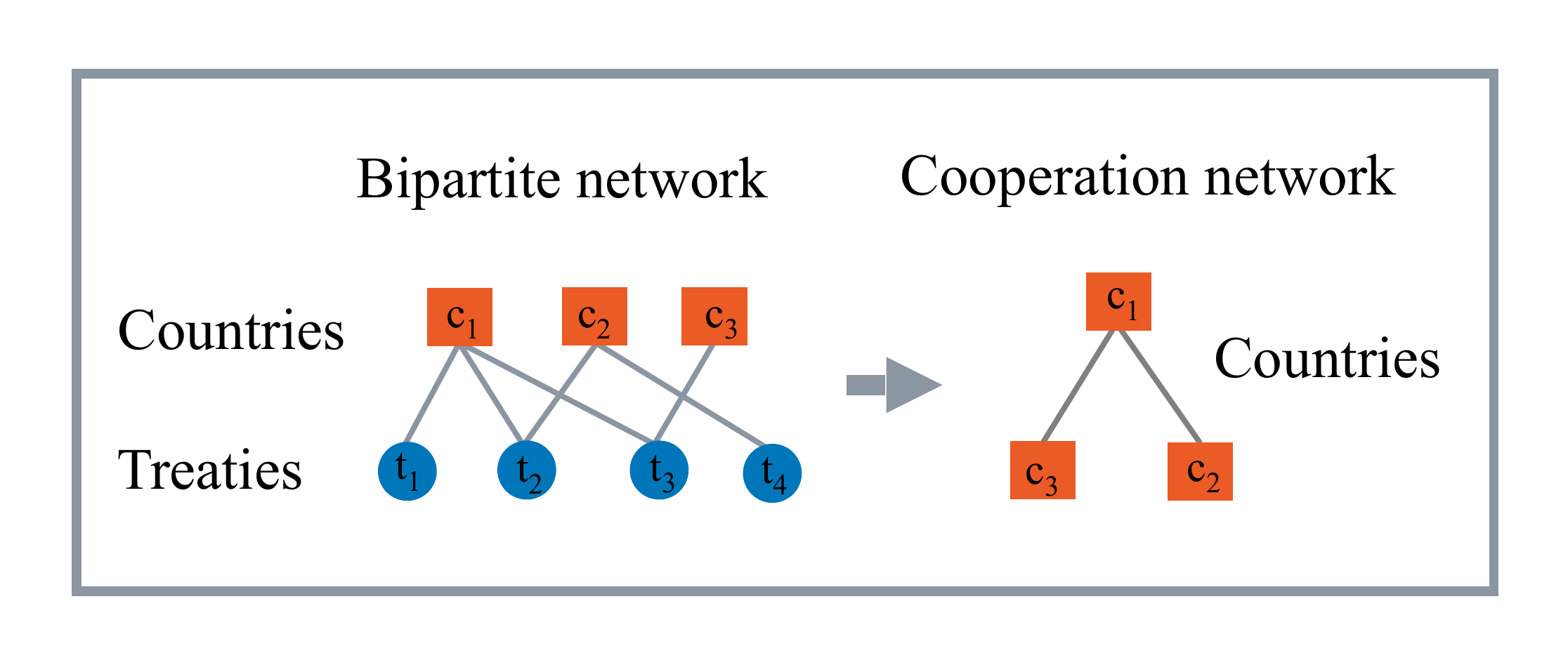}
 \caption{\textbf{Network construction}}
 \label{fig:construction}
 \end{figure}

In network analysis, affiliation data can be abstracted as a bipartite network. Also known as two-mode networks or affiliation networks, bipartite networks have two disjoint classes of nodes, participants and groups or events, and links connecting participants to groups or events~\citep{latapy2008basic}. Accordingly, we can represent the country-treaty relationships as a bipartite network in which, if a country signs a treaty, a link is created between the country and the treaty, as shown in the left-hand panel of Fig.~\ref{fig:construction}. 

\subsubsection{One-mode projections}

The cooperative tie between two countries is defined as the co-affiliation with, or co-participation in, the same treaties. In other words, if two countries are signatories of the same treaty or treaties, there is a cooperative tie between them, as shown in the right-hand panel of Fig.~\ref{fig:construction}. To obtain these cooperative ties, we need to project the bipartite network defined above onto a one-mode network using the country layer. In network science, this process is called one-mode projection and the resulting networks are called one-mode networks. We call the annual one-mode networks obtained through this process the environmental cooperation network, or cooperation network for short. 

Links are not just binary, i.e., either present or absent, but are characterised by their intensity or strength. Heterogeneity in the intensity of links encodes valuable network information~\citep{barrat2004architecture}. We quantify the intensity of cooperation by assigning a weight to each link, which is proportional to the number of treaties two countries have co-signed and inversely proportional to the number of signatory countries involved in each common treaty~\citep{newman2001scientificb}. This implies that, all else being equal, bilateral treaties contribute more to the intensity of cooperation between two countries than multilateral treaties.

\subsubsection{Bipartite null models and statistically validated projections}

To ensure that the cooperation network truly reflects the relationship between countries, we filter out any connections that might also be found in a random network where links are assigned by chance. That is, we  remove any links that are not statistically significant. A variety of methods have been proposed to determine which links are significant~\citep{serrano2009extracting, neal2014backbone, latapy2008basic, saracco2017inferring}. Here we adopt the grand canonical algorithm proposed by~\cite{saracco2017inferring}, which can be used to obtain a statistically-validated projection of any binary, undirected, bipartite network. The general idea underpinning this method is that any two countries should be connected in the corresponding one-mode projection, i.e., the cooperation network, if, and only if, they co-signed a statistically significant number of treaties. 

The algorithm can be applied through the following four steps. First, for each pair of countries, the number of co-signed treaties is computed. This can be regarded as a measure of the degree of similarity between the two countries. 

The second step quantifies the statistical significance of the similarity between each pair of countries. The null hypothesis here is that the observed similarity between any two countries can be explained simply by chance, given the involvement of the two countries in various treaties. To test this hypothesis, an appropriate null model is needed.\footnote{The Python code for this step can be obtained from  \url{https://github.com/tsakim/bipcm}.} Here, we adopt the bipartite partial configuration model. This model is part of the entropy-based exponential random graph (ERG) class of null models, and constrains only the degrees of the nodes in the layer of interest, i.e., in our case the number of treaties each country has signed ~\citep{saracco2017inferring,saracco2015randomizing,park2004statistical,squartini2011analytical}. 

More specifically, the partial configuration model generates a bipartite network in which each country has exactly the same degree (i.e., participation in the same number of treaties) as in the original bipartite network, but the connections between countries and treaties have been randomly reshuffled.
Given two countries $c_i$ and $c_j$, the distribution describing the behaviour of each value of similarity between $c_i$ and $c_j$ is the  Poisson–Binomial distribution. That is, the Poisson-Binomial distribution describes the probability that two given countries $c_i$ and $c_j$ co-sign $n^{T}_{c_i,c_j}$ treaties simply by chance, with $n^{T}_{c_i,c_j}=[0,..., N_T]$, and where $N_T$ is the total number of treaties. 
Based on this bipartite partial configuration model, measuring the statistical significance of the observed value $n'^{T}_{c_c,c_j}$ thus implies calculating a $p-$value$_{c_i,c_j}$ on the Poisson–Binomial distribution, i.e., the probability that $c_i$ and $c_j$ co-sign a number of treaties greater than, or equal to, the $n'^{T}_{c_c,c_j}$ simply by chance. Notice that, as a one-tail statistical test, this approach would lead to establishing a link between any two countries if the observed number of co-signed treaties is ``sufficiently large''.

Third, once the $M$ $p-$values associated to each pair of countries have been calculated (where $M={N^{C}\choose{2}}$ is the total number of possible pairs of countries and $N^{C}$ the total number of countries), we adopt a statistical procedure for simultaneously testing multiple hypotheses of similarities between pairs of countries. This is necessary to account for the lack of independence of similarities (and associated $p-$ values), since each observed link in the original bipartite network between a given country and a given treaty inevitably affects the number of common treaties that country co-signs with each of the remaining countries, and therefore the similarities of several pairs of countries. To account for this, we applied the so-called False Discovery Rate (FDR) procedure, which controls for the expected number of false ``discoveries'' (i.e., incorrectly-rejected null hypotheses,~\citealt{benjamini}). To this end, we sort the $M$ $p-$values in increasing order and then identified the largest integer $\hat{i}$ such that $p-\textnormal{value}_{\hat{i}} \leq \frac{\hat{i}\alpha} {M}$, where $\alpha$ is the single-test significance level, which here we set at $0.01$. 

As a final fourth step, we obtain a statistically-validated projection of the bipartite network by considering as statistically significantly similar only those pairs of countries $c_i$ and $c_j$ whose $p-\textnormal{value}_{c_i,c_j} \leq p-\textnormal{value}_{\hat{i}}$. Equivalently, this translates into rejecting the null hypotheses of observing by chance the similarities between countries when the corresponding $p-$values are smaller than the given FDR threshold. In this way, a link will be established only between pairs of countries that are sufficiently similar, i.e., that have co-signed a larger number of treaties than would be randomly expected. All our subsequent analysis will be based on such a statistically validated network projection. 

\subsection{Network analysis}

We adopt global metrics from network science to quantify the topological structure of the environmental cooperation network as it evolved over time. Our chosen metrics include measures of network size (cumulative frequency of nodes and links), connectivity (average degree, average strength), and social cohesion (density, shortest path length, number of components, and clustering coefficient). In addition, the roles of countries in the cooperation network are investigated through centrality measures, such as betweenness centrality and closeness centrality.

\textbf{Cumulative frequency of nodes and links.} The size of a network can be measured straightforwardly through the number of nodes and links it contains. In a dynamic setting the growth in network size can be measured through cumulative distributions of nodes and links over time.  

\textbf{Degree and strength.} The degree $k$ of a node is the number of links connected to it. In weighted networks, the metric of node degree is complemented by node strength, $s$, which is the sum of the weights of the links incident upon the node~\citep{barrat2004architecture}. In our cooperation network, the degree of a country indicates the number of partners which this country cooperates with, while the strength accounts for the intensity of cooperation between this country and others. The average degree and average strength of a network are global variables of network connectivity. In contrast, the degree and strength of individual nodes are local measures of connectivity. A node with a higher degree is expected to have more access to information and to be more salient for communication activities in the network than nodes with lower degrees~\citep{hafner2009network, freeman1978centrality}.

\textbf{Density.} The density of a network is the ratio between the actual number of links $m$ and the maximum possible number of links, i.e., $\binom{n}{2}=\frac{1}{2}n(n-1)$, where $n$ is the number of nodes in the network. Density ranges from $0$, when no link is established, to $1$, when all possible links have been established. In the cooperation network, density measures the portion of the potential cooperative connections that are actual connections through treaties. Thus, the network density can be seen as an indicator of cooperative cohesion among countries.

\textbf{Shortest path length.} For a binary network, the shortest path length $d_{ij}$ between node $i$ and node $j$ is the length of the path with the lowest number of links separating the two nodes~\citep{newman2018networks}. In weighted networks, shortest path lengths between nodes are traditionally measured through the algorithm proposed by~\cite{dijkstra1959note}. In this case, weights indicate the cost of information transmission or resource flow, and distances are calculated as sums of the weights of the links traversed. Thus, the weighted shortest path length between any two nodes is the path with the least resistance in terms of exchange costs. However, in our study the weights of links do not represent the cost, but the intensity of cooperation between countries, and therefore we use the reciprocal of weights to identify weighted shorted paths using the Dijkstra’s algorithm~\citep{newman2001scientificb,brandes2001faster}. Hence, in our network, the higher the weight of the link, the closer two countries are and the lower the cost of cooperation. 

\textbf{Component.} A component is a largest subset of nodes in a network in which there exists at least one path between any pair of nodes. The components in a network organise the network into different isolated subgraphs, and the number of components in a network can therefore be used to assess isolation of nodes. All else being equal, a network with more (and smaller) components is less cohesive, as countries only build cooperative ties within the same component. Conversely, a smaller number of (larger) components in the cooperation network indicates a higher level of network cohesion.

\textbf{Clustering coefficient.} 
Studies of the network sources of social capital have suggested that closed structures facilitate access to complex information, stimulate trust, sustain cooperation and promote social norms by enabling the enforcement of collective sanctions~\citep{burt2000network,coleman1988social}. Traditionally, network closure is measured through the global and local clustering coefficients. 

The global clustering coefficient of a network quantifies the level of \emph{global connectivity} based on density of triplets. A triplet can be defined as three nodes connected by either two (open triplet) or three (closed triplet) links. The global clustering coefficient measures the fraction of closed triplets over the total number of open and closed triplets, that is, the degree to which triplets in a network close up into triangles~\citep{newman2018networks,opsahl2009clustering}. For example, in the context of international relations, it has been shown that countries that share bilateral agreements with the same third parties are more likely to form bilateral agreements themselves ~\citep{kinne2013network}.

To take the weights of links into consideration, we use a generalisation of the global clustering coefficient based on the values of triplets ~\cite[]{opsahl2009clustering}:

\begin{equation}
C^w=\frac{\sum_{\text {closed triplets}} v_{i}}{\sum v_{i}}
\end{equation}
 
Here, the value of a triplet $v_i$ is the arithmetic mean of the weights of the two links that make up the triplet. Note that the weight of the closing link of a triplet is not taken into account as the weighted coefficient is simply aimed at assessing the likelihood of the closing link, and not its strength.

Unlike the global clustering coefficient, the \emph{local clustering coefficient} is defined for a single node, and captures the connectivity of a node's local neighbourhood. In particular, it quantifies the tendency of a node's neighbours to be connected with each other. 

 The \emph{weighted local clustering coefficient} is a generalisation of the coefficient that takes the weights of links into consideration~\citep[see][for details on comparison of different methods]{saramaki2007generalizations}. We rely on the method proposed by~\cite{onnela2005intensity} to account for the intensity of cooperation between countries. This method is based on a node's subgraph intensity, defined as the geometric average of the weights of the links forming all closed triplets centred on the node, where each weight is normalised by the maximum weight globally found in the network.

In addition, in what follows we discuss findings based on an alternative method proposed by~\cite{barrat2004architecture}, according to which the contribution of each closed triplet centred on a node depends on the ratio of the average weight of the two links incident on the node to the average strength of the node (i.e., the node's strength divided by the node's degree). Hence, in this case local distributions of weights heavily affect the value of the weighted local clustering coefficient, % according to \cite{barrat2004architecture}, 
 while according to the former method proposed by \cite{onnela2005intensity} the coefficient depends on the distribution of weights across the whole network. 

\textbf{Betweenness centrality.} Betweenness centrality was originally proposed by~\cite{freeman1977set} to measure the degree to which one node lies on the shortest paths between others. It is defined as: 

\begin{equation}
C_{B,i}=\sum_{j,k}\frac{g_{j,k}^i}{g_{j,k}}, \label{betweenness}
\end{equation}

\noindent where $g_{j,k}$ is the number of shortest paths between node $j$ and node $k$, and $g_{j,k}^i$ is the number of those paths passing through node $i$. If $j=k$, $g_{j,k}=1$, and if $i \in j,k $, then $ g_{j,k}^i=0$. 

Betweenness centrality quantifies the extent to which a node presides over indirect connections between all other nodes in a network~\citep{burt2000network}. Hence, betweenness centrality is an indicator of the importance of nodes in participating in, and controlling, the flow of critical resources in networks, such as the the spread of information, news, opportunities across various regions of a social system~\citep{freeman1978centrality}. In international relations networks, a node with a high betweenness centrality has a high brokerage power over otherwise disconnected countries, and has therefore the potential to foster and facilitate cooperation between other countries~\citep{hafner2009network}.

\textbf{Closeness centrality.} The \emph{closeness centrality} of a node is defined as the inverse of the average shortest path length from the node to all other reachable nodes: 

\begin{equation}
C_{C,i}=\frac{n-1}{\sum_{j}d_{ij}},\label{closeness}
\end{equation}

\noindent where $n$ is the number of nodes reachable by node $i$, and $d_{ij}$ is the shortest path length between node $i$ and node $j$. In social networks, higher closeness centrality, i.e., shorter average distance from other nodes, implies quicker communication at a lower cost~\citep{freeman1978centrality}. Information from the most central nodes can spread out quickly and in the most cost-effective manner. Thus, in our study closeness centrality can be a proxy of the proximity of a country to other countries in the network based on existing cooperative connections, and consequently of the potential cost for sustaining cooperation with other countries.

\section{The growth in environmental treaties}\label{Sec:motivation}

Before turning to network analysis, we provide  a brief quantitative description of the raw country-treaty-year data on international environmental agreements (IEAs). 

Over the past decades the number of IEAs has grown significantly. Countries have become more active in joining them and the range of topics they cover has increased \citep{mitchell2020we,mitchell2003international}. We can summarise these trends through three statistics: the number of signatories per treaty, the number of treaties signed by each country, and the number of environmental issues covered by IEAs. 

\begin{figure}[H]
\centering
\begin{subfigure}[c]{0.46\textwidth}
\centering
\includegraphics[width=\textwidth]{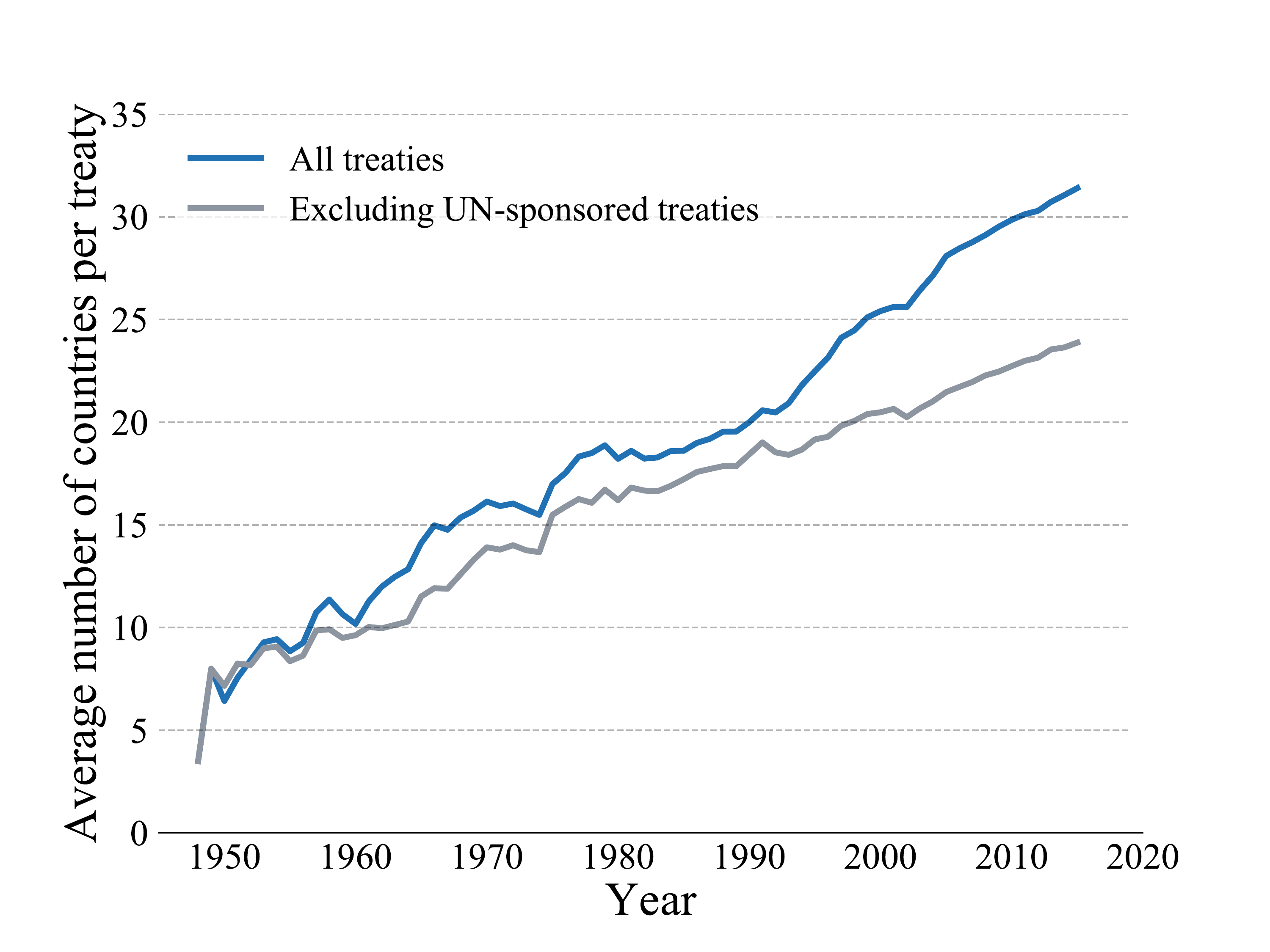}
\caption{Average number of countries per treaty}
\label{fig:average_treaty_degree_Bi}
\end{subfigure}
\begin{subfigure}[c]{0.46\textwidth}
\centering
\includegraphics[width=\textwidth]{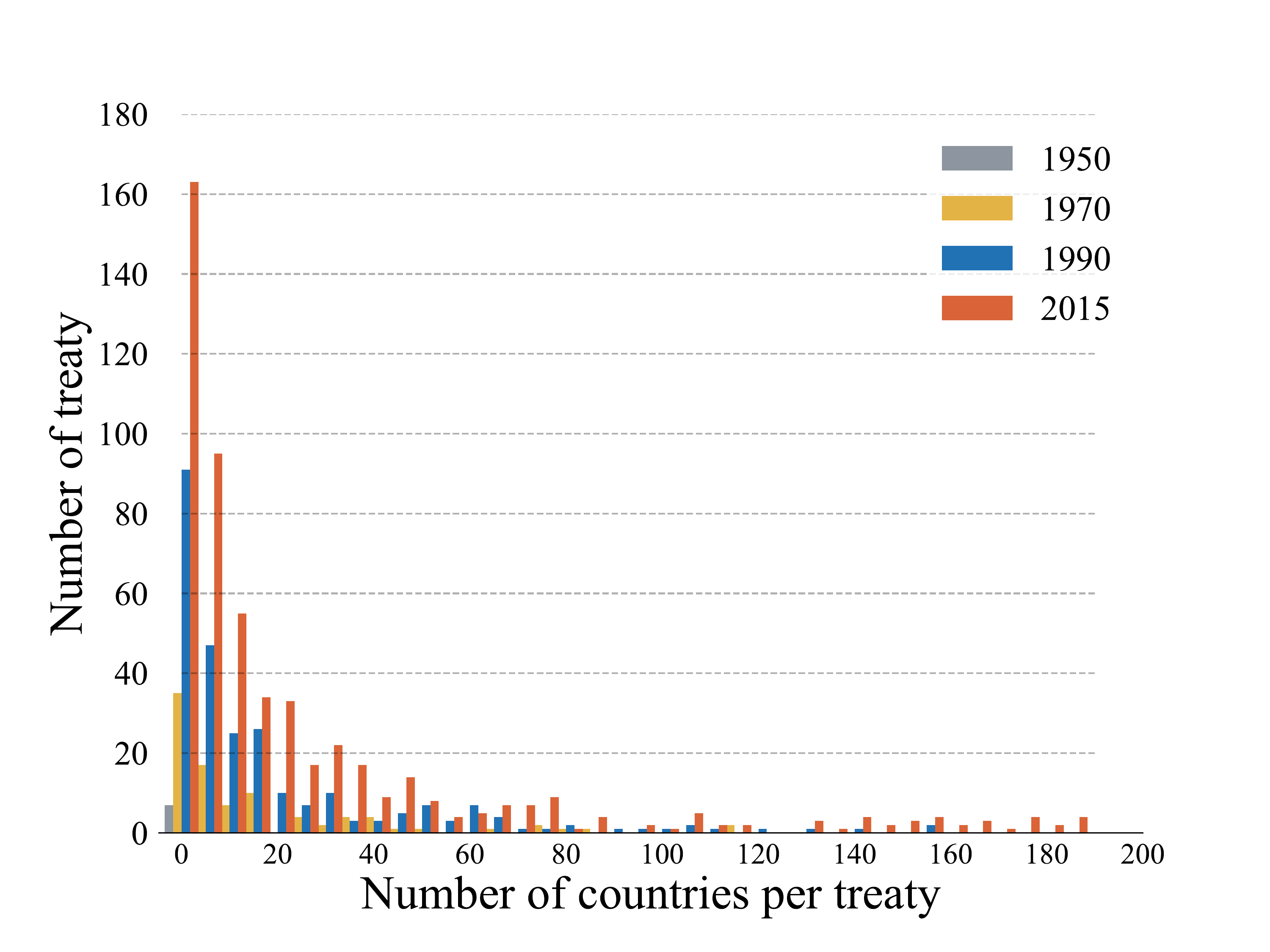}
\caption{Distribution of the number of countries per treaty in different years}
\label{fig:treaty_degree_Bi}
\end{subfigure}
\vspace*{3mm}
\caption{Number of countries per treaty.}
\label{fig:treaty_layer_degree}
\end{figure}

Over the period under scrutiny, the average number of signatory countries per treaty rose from $4$ in 1948 to $31$ in 2015 (Fig.~\ref{fig:treaty_layer_degree}, panel a). At the same time, the distribution of the number of signatories per treaty has become wider and more skewed (panel b). We have seen the emergence of global treaties that are signed by a large number of countries (\textgreater 75 countries),\footnote{The ten largest treaties by number of signatories, in decreasing order of size, are: Vienna Convention for the Protection of the Ozone Layer, Montreal Protocol on Substances that Deplete the Ozone Layer,
Convention on Biological Diversity,
United Nations Framework Convention on Climate Change,
United Nations Convention to Combat Desertification in those Countries Experiencing Serious Drought and/or Desertification, particular in Africa,
Convention on the Prohibition of the Development, Production, Stockpiling and Use of Chemical Weapons and on their Destruction,
 Kyoto Protocol to the United Nations Framework Convention on Climate Change,
United Nations Convention against Transnational Organised Crime,
Basel Convention on the Control of Transboundary Movements of Hazardous Wastes and their Disposal,
WHO Framework Convention on Tobacco Control (FCTC).} but also a significant increase in the number of treaties with fewer than 10 signatories, suggesting that formal cooperation on both regional and global dilemmas has expanded over time.

\begin{figure}[H]
\centering
\begin{subfigure}[c]{0.46\textwidth}
\centering
\includegraphics[width=\textwidth]{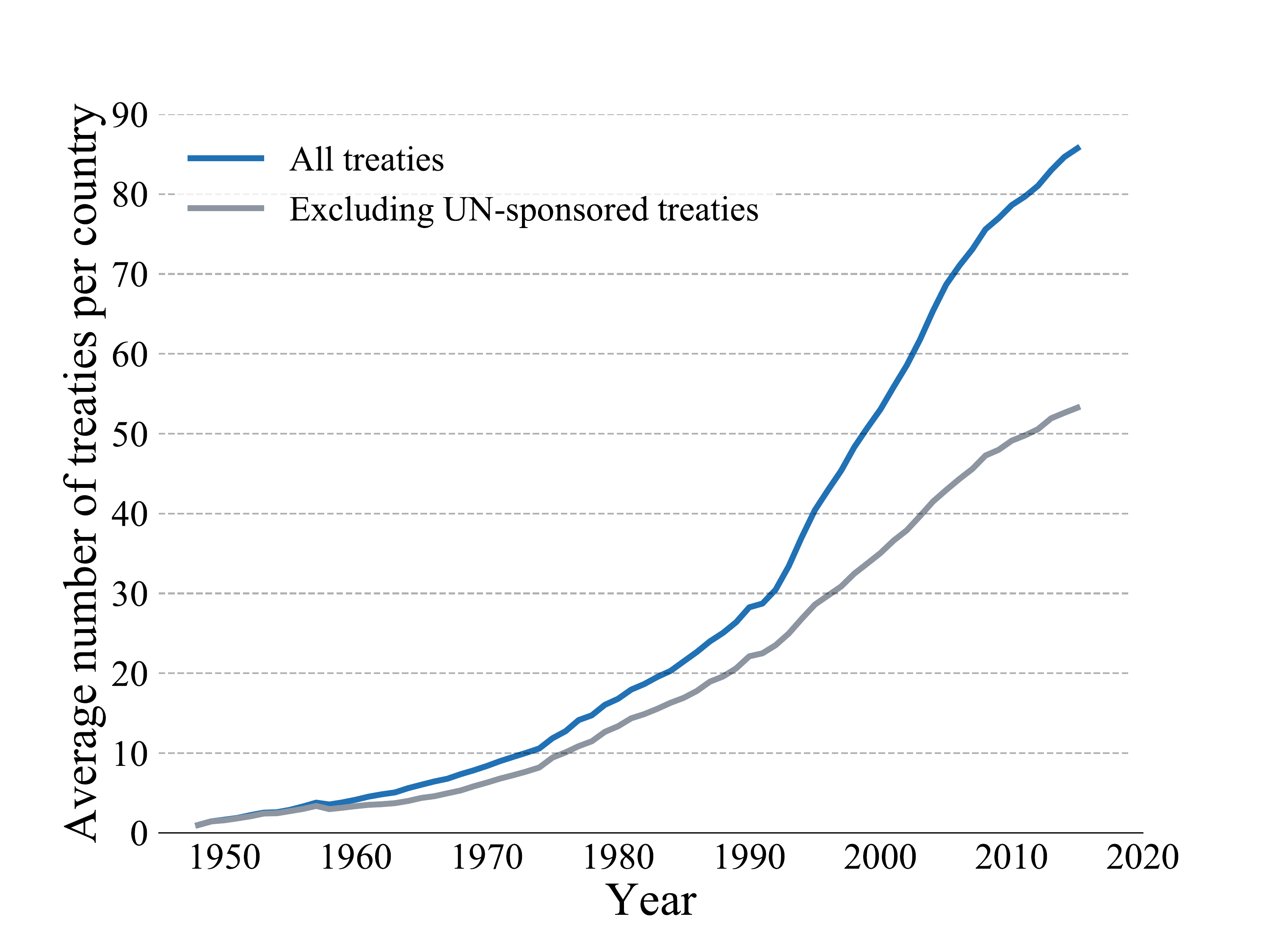}
\caption{Average number of treaties per country}
\label{fig:average_country_degree_Bi}
\end{subfigure}
\begin{subfigure}[c]{0.46\textwidth}
\centering
\includegraphics[width=\textwidth]{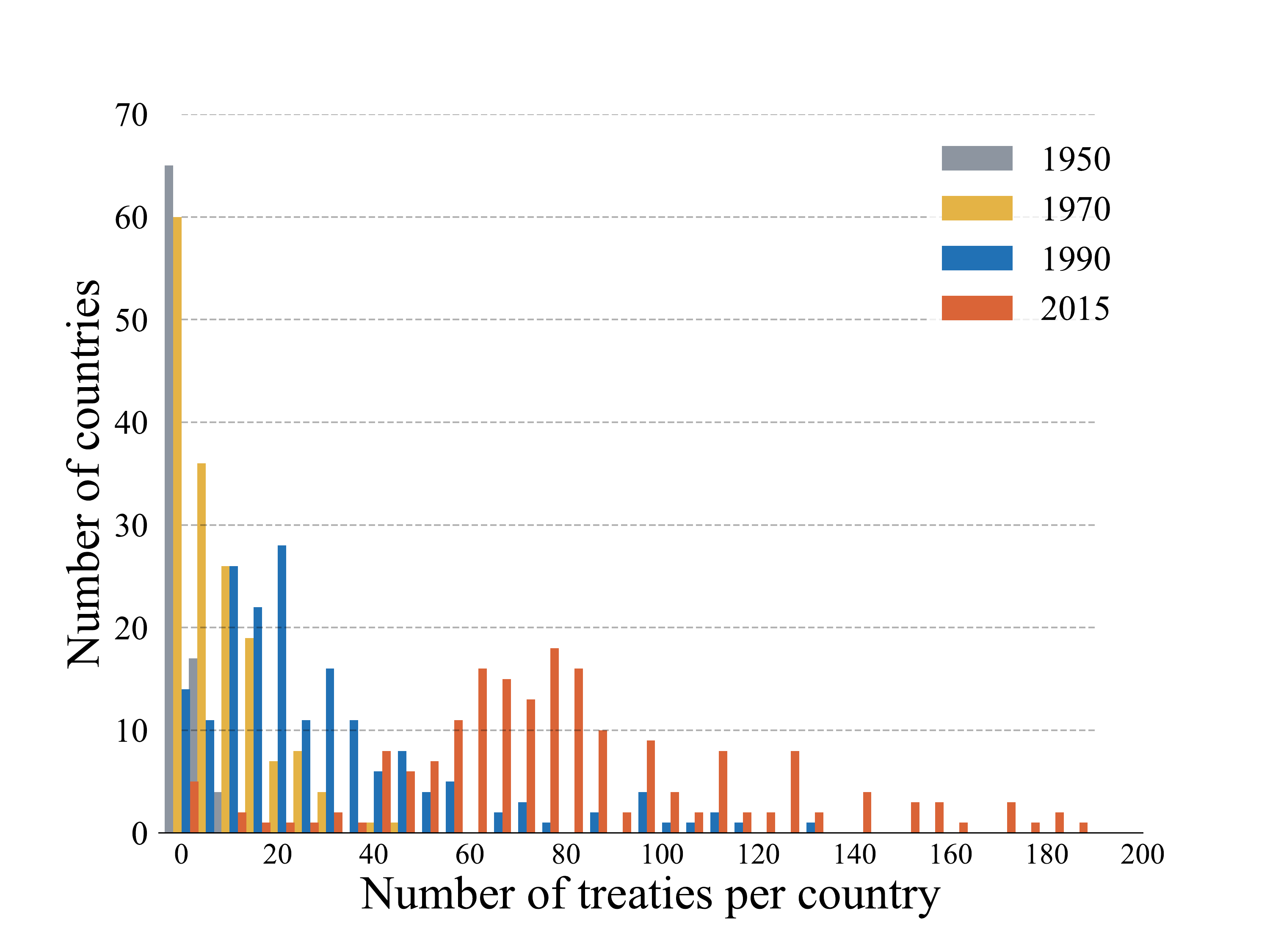}
\caption{Distribution of the number of treaties per country in different years}
\label{fig:country_degree_Bi}
\end{subfigure}
\vspace*{3mm}
\caption{Number of treaties per country.}
\label{fig:country_layer_degree}
\end{figure}

The number of treaties each country signs up to has gone up in parallel (Fig.~\ref{fig:country_layer_degree}). Growth was particularly fast between 1992 and 2008, when the average number of treaties per country grew from $30$ to $76$ (panel a). The average patterns mask some interesting heterogeneity (panel b).  In the first part of the period under analysis, most countries tended to join only a small number of treaties, while a small number of very active countries signed up to a large number. Over time, the distribution becomes less skewed. The absolute number of treaties increases, but the peak decreases and moves to the right. In 1950, a few leading countries (France, the Netherlands, the United Kingdom and the US) had signed over $10$ treaties. In 1970 the lead group had expanded to also include Belgium, Denmark, Switzerland and Sweden, each signing over $30$ treaties. In 1990, a larger group of still mostly European countries had signed more than $90$ treaties each and in 2015 they were signatories to over $190$ treaties each. 

\begin{figure}[H]
\centering
\includegraphics[scale=0.4]{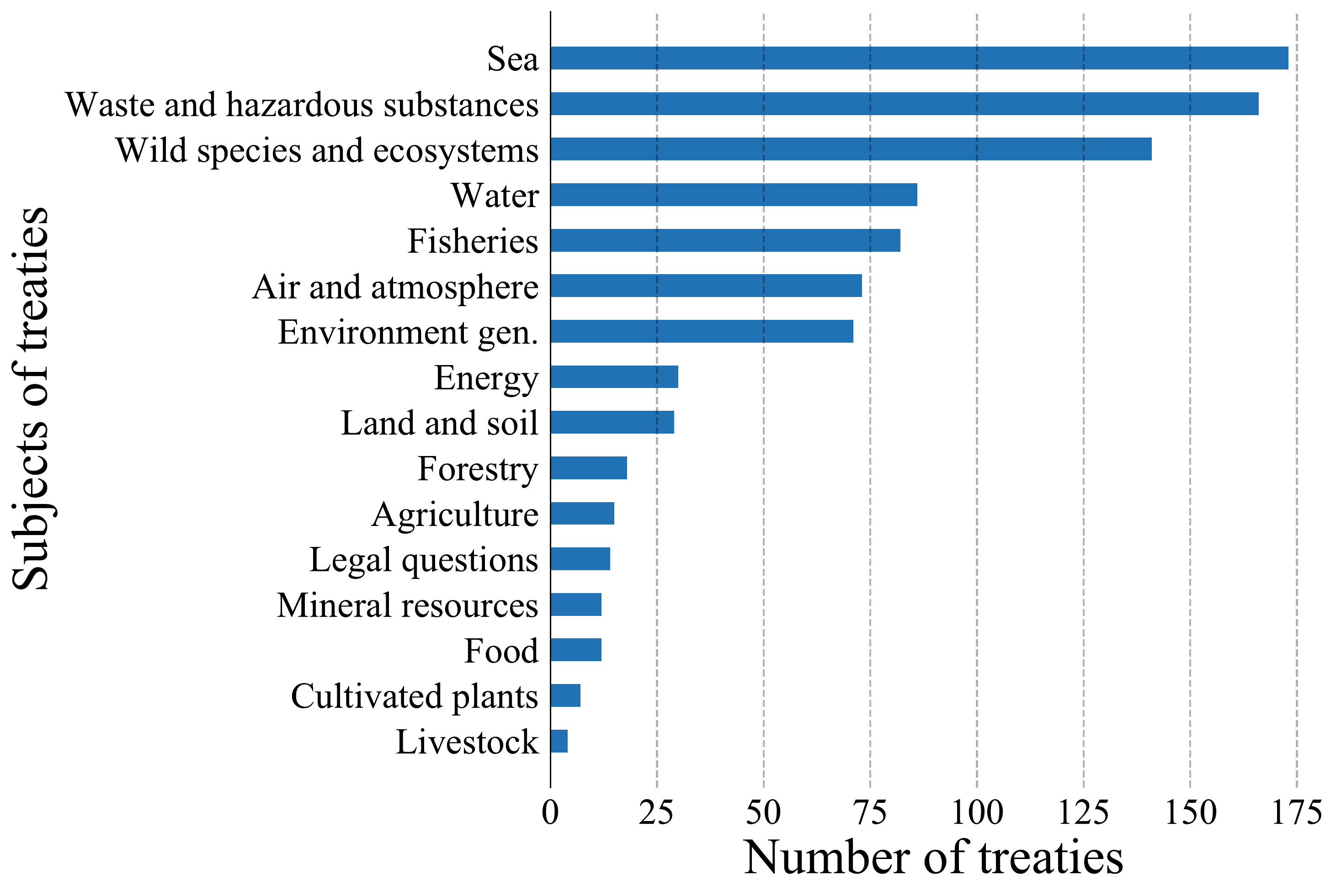}
\caption{Cumulative frequency of treaties for different subjects in 2015}
\label{fig:treaty_subjects_initial}
\end{figure}

By the year 2015, IEAs covered practically all aspects of regional or global environmental concern, as shown in Fig.~\ref{fig:treaty_subjects_initial}. There is considerable thematic overlap, with many treaties covering more than one subject area. For example a large number of treaties on the seas also concern issues of waste (57 treaties), fisheries (38 treaties) or wild species and ecosystems (17 treaties). In what follows, we aggregate IEAs into six categories: sea and fisheries, wild species and ecosystems, waste and hazardous substances, natural resources (e.g., water, cultivated plants, environment genes, food, forestry, land and soil, livestock, and mineral resources), air and atmosphere (e.g., air pollution, ozone layer depletion and climate change), and energy.

 21. 

\begin{figure}[H]
\centering
\begin{subfigure}[c]{0.3\textwidth}
\centering
\includegraphics[width=\textwidth]{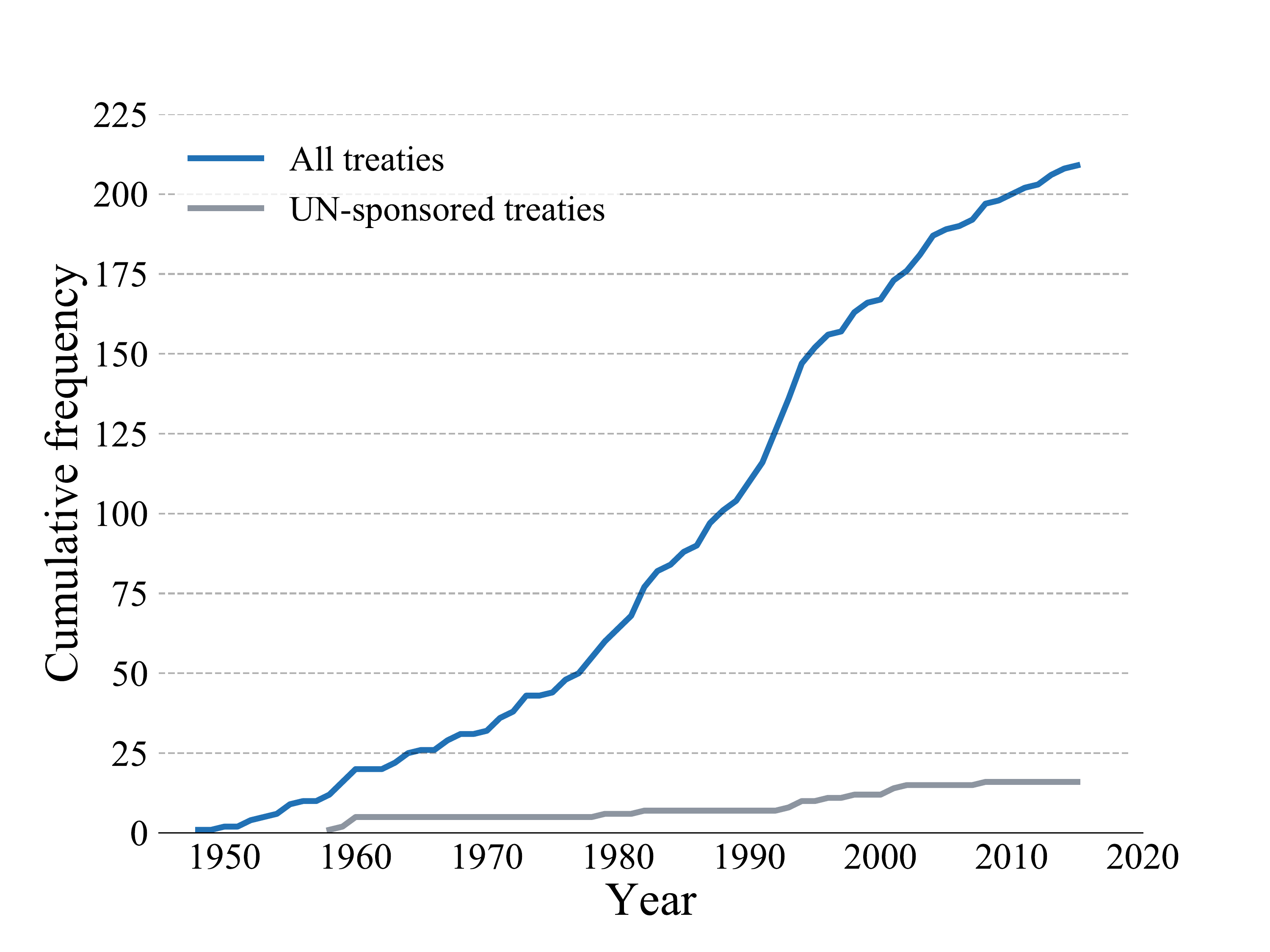}
\caption{Sea and fisheries}
\label{fig:num_sea}
\end{subfigure}
\begin{subfigure}[c]{0.3\textwidth}
\centering
\includegraphics[width=\textwidth]{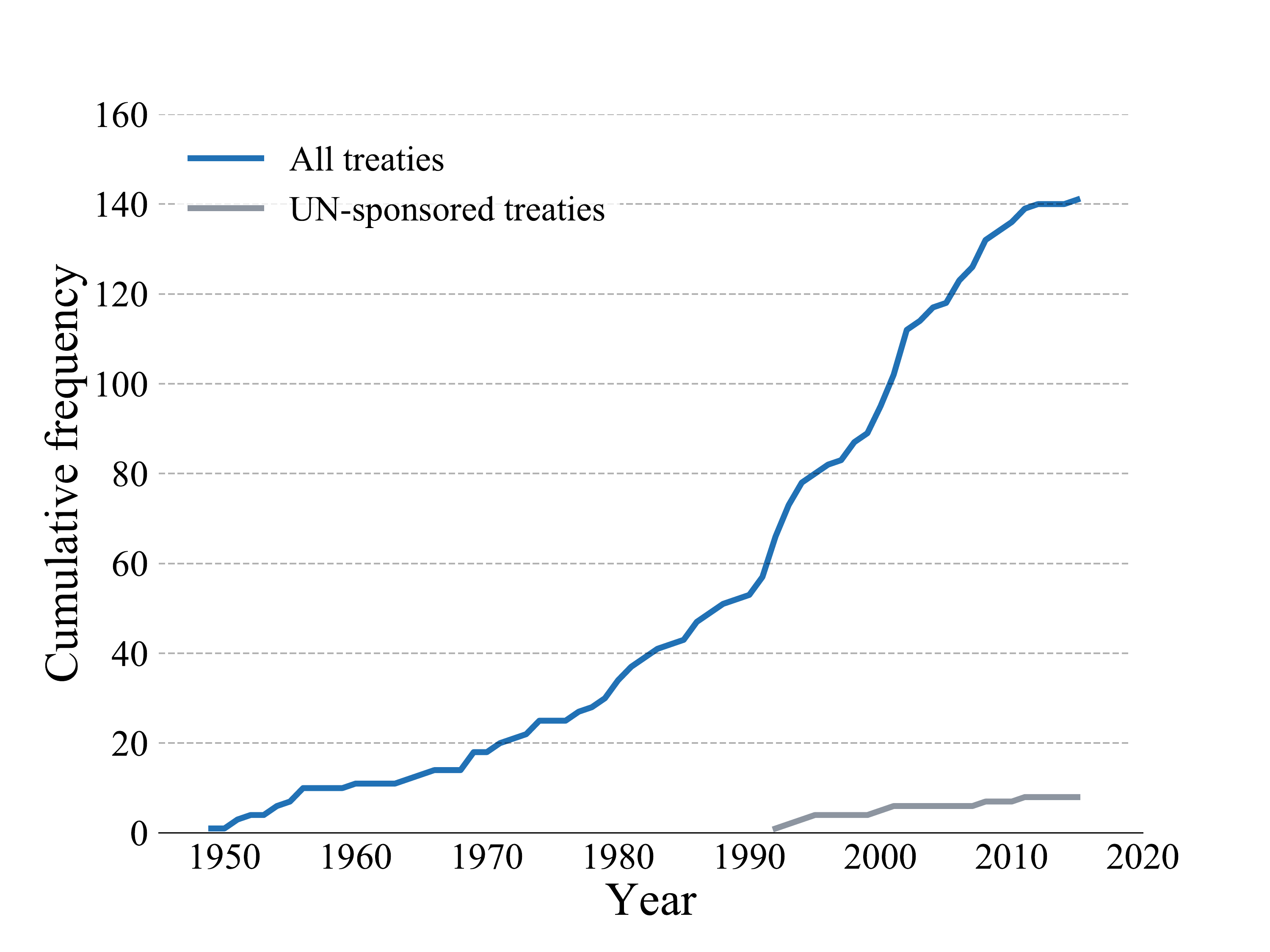}
\caption{Wild species and ecosystems}
\label{fig:num_species}
\end{subfigure}
\begin{subfigure}[c]{0.3\textwidth}
\centering
\includegraphics[width=\textwidth]{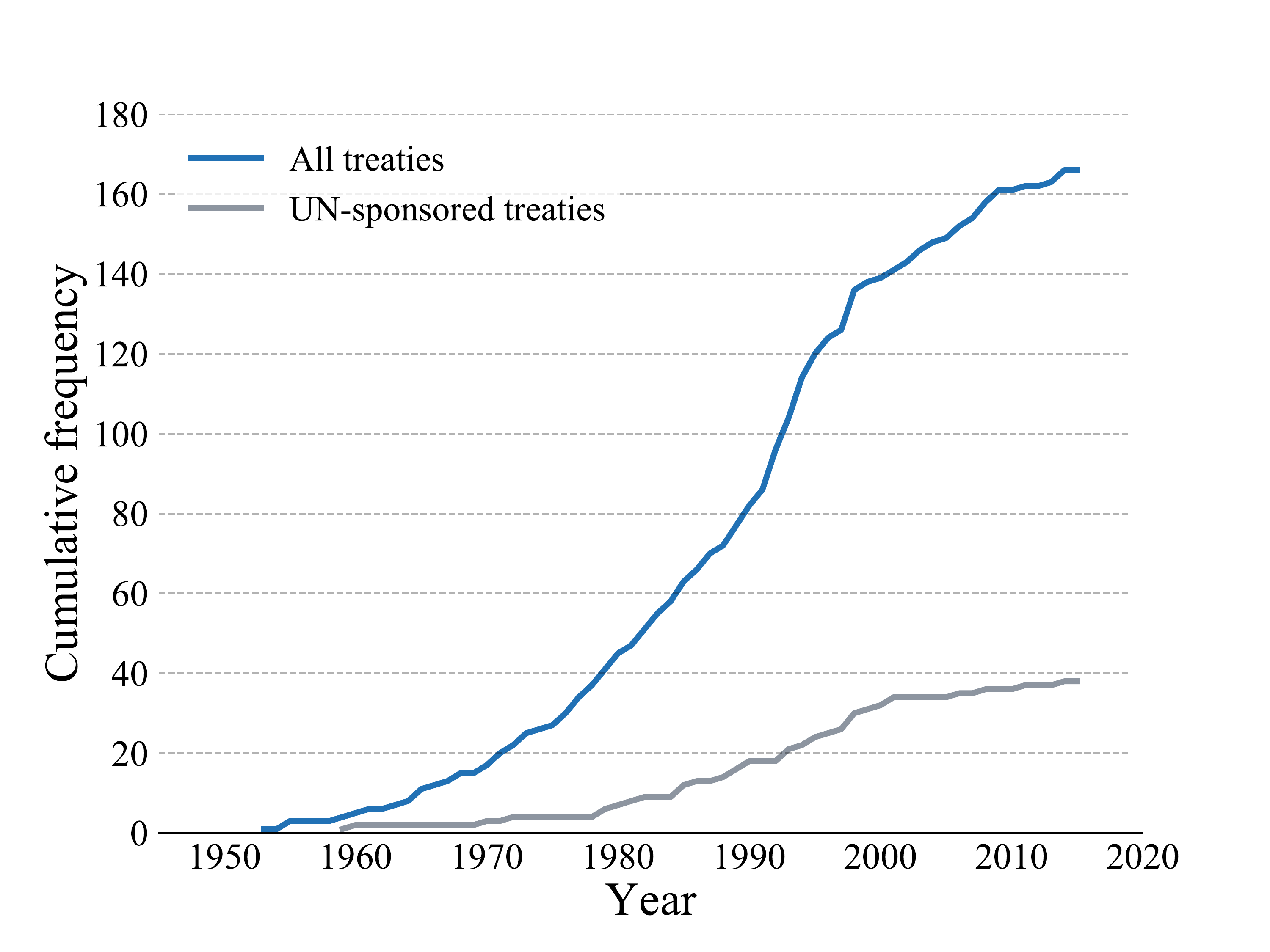}
\caption{Waste and hazardous substances}
\label{fig:num_waste}
\end{subfigure} \\
\begin{subfigure}[c]{0.3\textwidth}
\centering
\includegraphics[width=\textwidth]{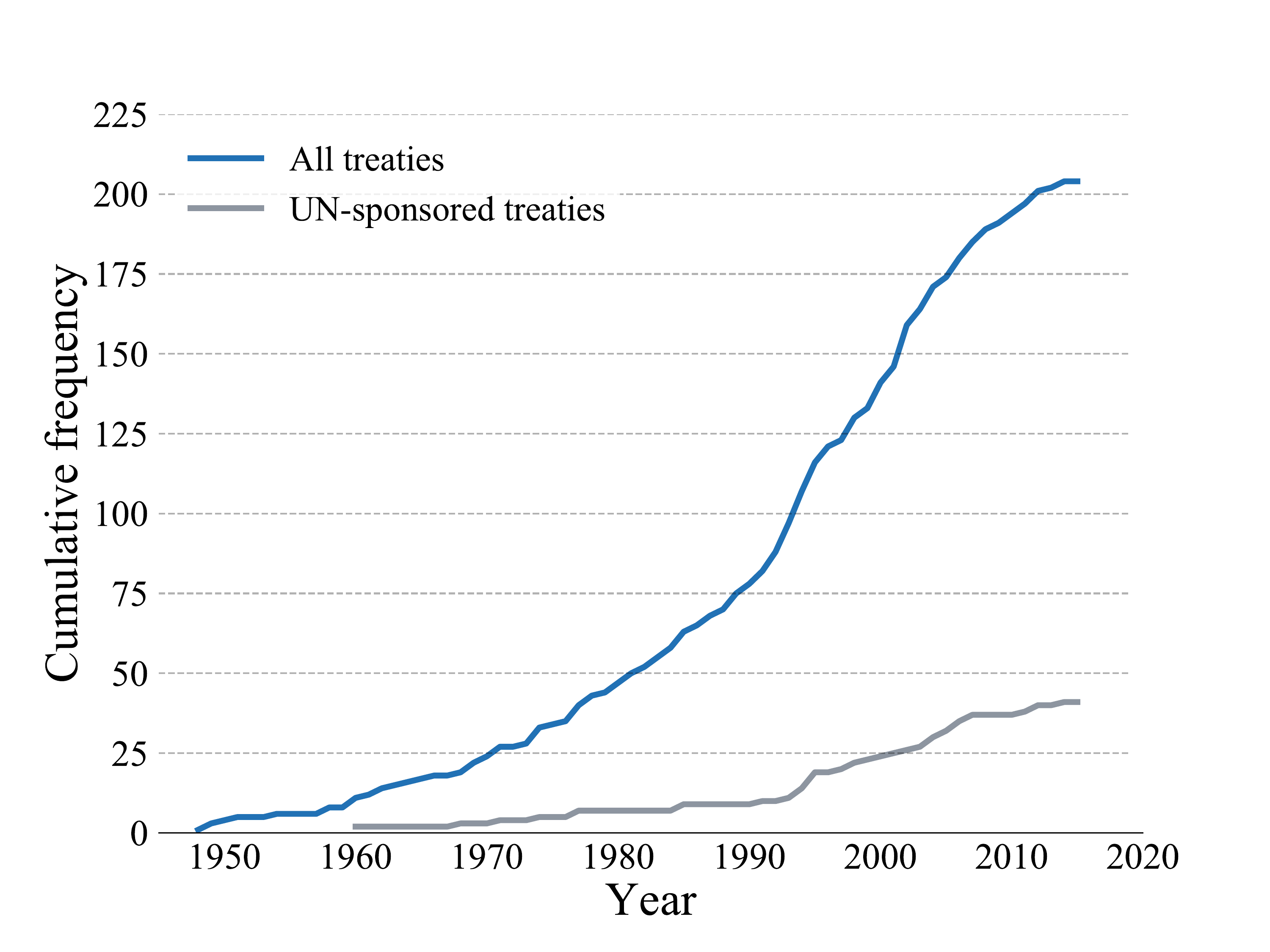}
\caption{Natural resources}
\label{fig:num_resource}
\end{subfigure}\begin{subfigure}[c]{0.3\textwidth}
\centering
\includegraphics[width=\textwidth]{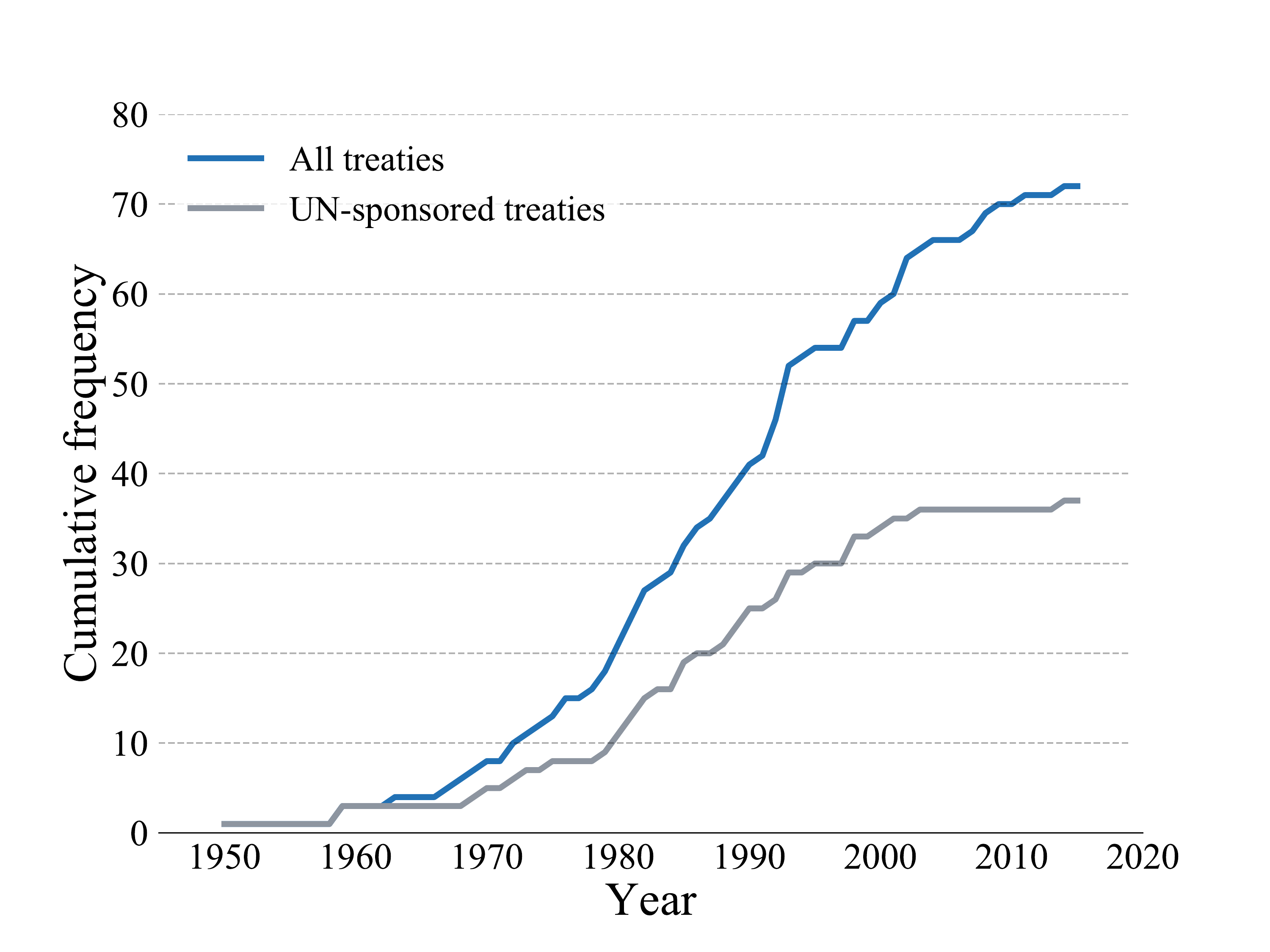}
\caption{Air and atmosphere}
\label{fig:num_air}
\end{subfigure}
\begin{subfigure}[c]{0.3\textwidth}
\centering
\includegraphics[width=\textwidth]{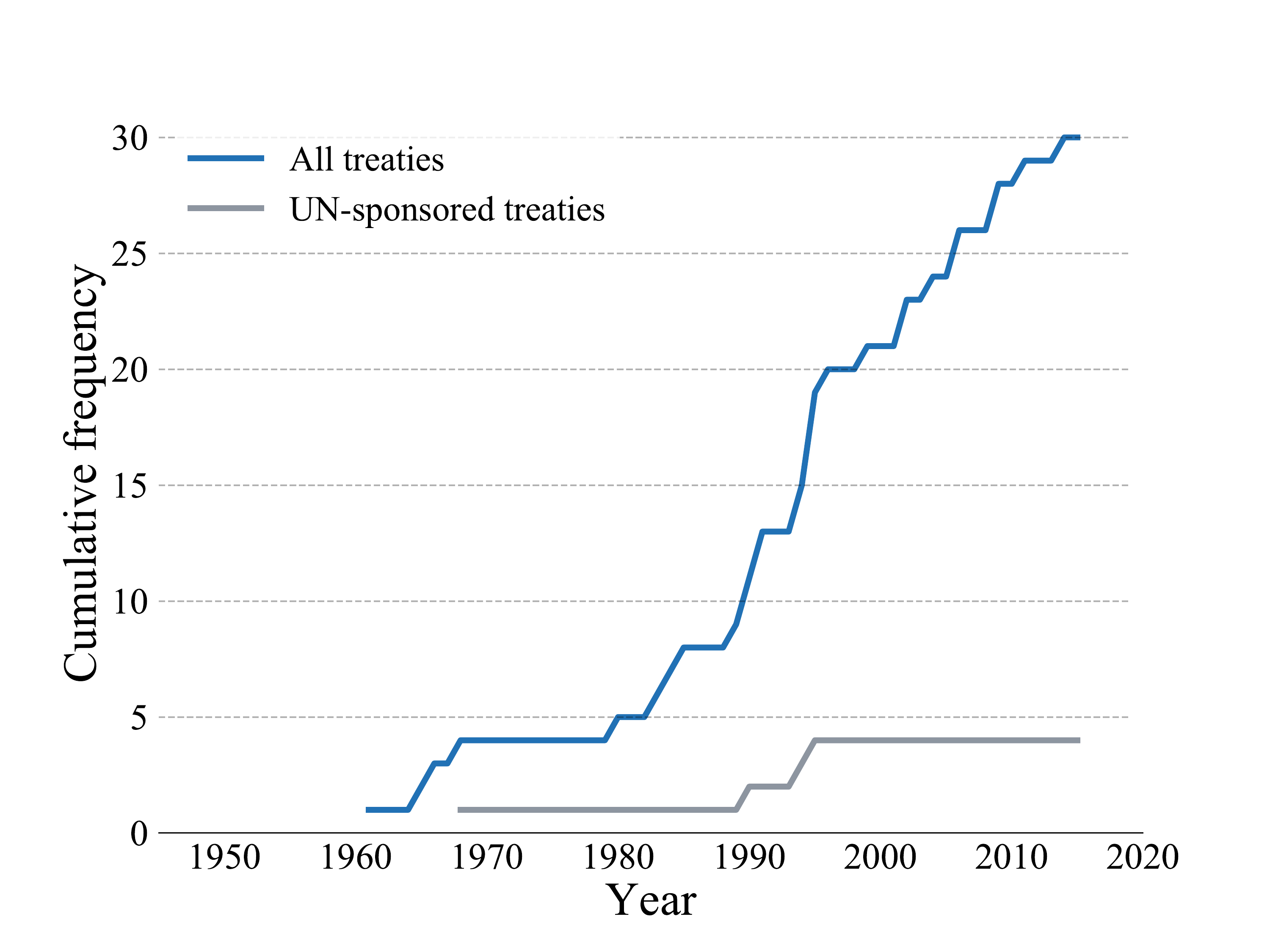}
\caption{Energy}
\label{fig:num_energy}
\end{subfigure}
\vspace*{3mm}
\caption{Cumulative frequency of treaties for different subjects}
\label{fig:num_subjects}
\end{figure}

The United Nations, through the UN Environment Programme and other agencies, have played a prominent role in facilitating these trends. Since it was established in 1945, the UN has promoted over 100 IEAs. However, Figs~\ref{fig:treaty_layer_degree}, ~\ref{fig:country_layer_degree}, and ~\ref{fig:num_subjects} suggest that the broader pattern of increased environmental cooperation is not driven primarily by the UN. A crucial exception is IEAs on air and atmosphere, where UN treaties account for about $50\%$ of the total, compared with less than $25\%$ in all other categories. 

Together, the above statistics describe how the global system of environmental governance has become larger, more inclusive and more comprehensive over time. To understand the systemic implications of these trends we turn to network analysis.

\section{The extent of cooperation}\label{sec:size_and_connectivity}
\subsection{Overview}
We first explore what the growth in IEAs means for the strength and depth of environmental collaboration. Intuitively, one would expect the proliferation of treaties documented in Section \ref{Sec:motivation} to result in deeper and more intensive environmental cooperation. Our interest is in the topological properties of the cooperation network and when statistically meaningful environmental cooperation emerged.

The first metrics we turn to concern network size and connectivity. A straightforward way to measure the size of the environmental cooperation network is the number of nodes (countries) and links (through treaties) it contains, and more specifically the cumulative frequency of nodes and links over time. We use two metrics to measure the connectivity of the network, i.e., the average degree and the average strength. Recall from section \ref {Sec-methodology} that the average degree considers the average number of partners with which each country cooperates, while the average strength describes the average intensity of cooperation of a country with others.

This analysis reveals a first stylised fact about the international network of environmental cooperation. 
\\
\textbf{Stylised Fact 1:} \textit{ Meaningful environmental cooperation started in the early 1970s, and since then countries have been integrated into a network of increasingly intensive environmental cooperation. The growing intensity of global environmental cooperation is reflected in the size of the network, which includes virtually all countries of the world, and a high level of connectivity (high average degree and node strength) between countries. The UN has been an important platform for, but not the main contributor to, the connectedness of the environmental cooperation network.}  

\subsection{Network size}
A first important observation when assessing the size of the environmental cooperation network is that a statistically significant network only appeared in 1971. From 1948 to 1970, the number of common treaties between any two countries is not significantly different from the random connections in the null model. 

\begin{figure}[H]
\centering
\begin{subfigure}[c]{0.46\textwidth}
\centering
\includegraphics[width=\textwidth]{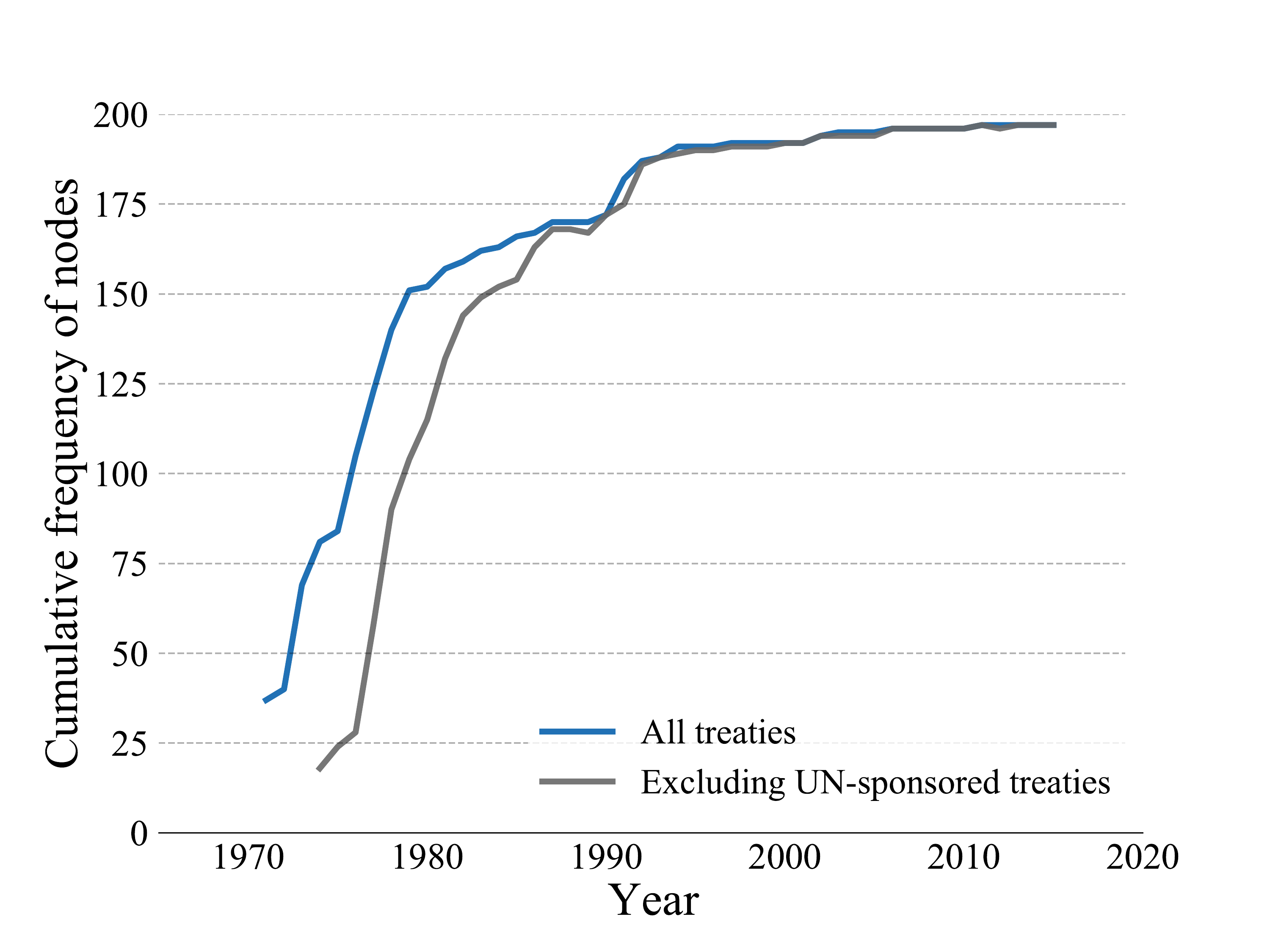}
\caption{Cumulative frequency of nodes}
\label{fig:nodes_country}
\end{subfigure}
\begin{subfigure}[c]{0.46\textwidth}
\centering
\includegraphics[width=\textwidth]{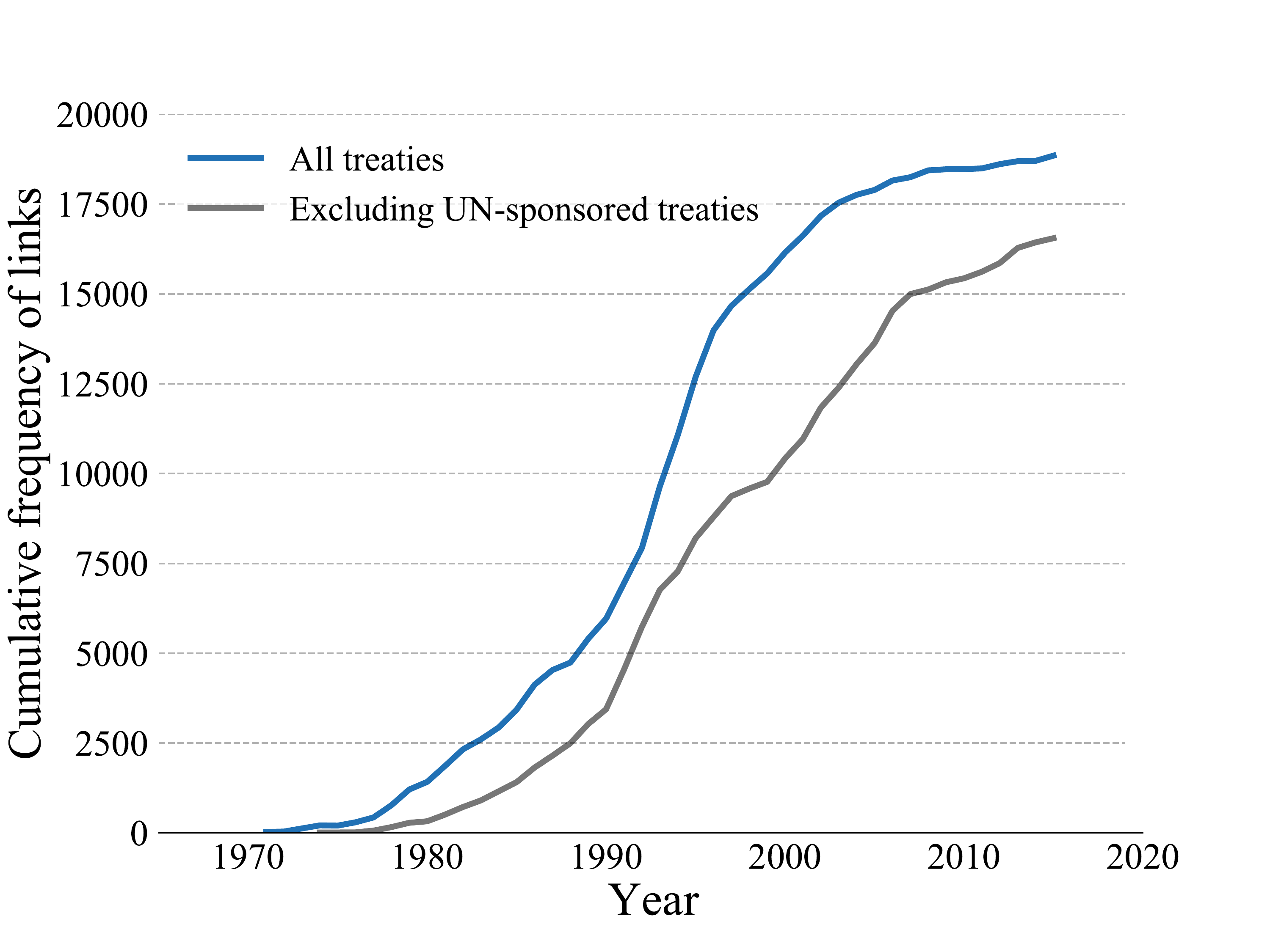}
\caption{Cumulative frequency of links}
\label{fig:links_country}
\end{subfigure}
\vspace*{3mm}
\caption{Cumulative frequency of nodes and links in country networks from 1971 to 2015.}
\label{fig:nodes_links_country}
\end{figure}

However, since then the cumulative frequency of network nodes and network links has grown steadily, as shown in Fig.~\ref{fig:nodes_links_country}. The number of participating countries (network nodes) grew particularly fast in the early 1970s, when many of the newly independent countries in the Global South began to engage with the international environmental treaties. The cooperation network became stable in the year 1980, when the number of new countries in the network begins to level off. 

The emergence of a stable, statistically significant cooperation network in the early 1970s corroborates the view of many international relations scholars, who see the advent of the 1972 UN Conference on the Human Environment in Stockholm as the beginning of a systematic and potentially universal approach to international environmental policy-making \citep{falkner2019emergence}.

The most rapid growth in network links occurred in the 1990s. During this period, $153$ treaties promoted cooperative ties among $192$ countries. The growth rate in both nodes and links levelled off around the year 2000, when nearly all countries were members of the cooperation network and the cumulative frequency of links almost reached its maximum.

We also investigated the role of the United Nations as a platform for international  environmental cooperation. The fact that the cooperation network became statistically significant in the advent of the 1972 Stockholm conference and its rapid growth after the 1992 UN "Earth Summit" in Rio de Janeiro suggests that UN treaties played an important role in encouraging countries to engage on global environmental issues. The suite of treaties agreed in Rio have come to define global environmental cooperation in areas such as biodiversity (Convention on Biodiversity), climate change (UN Framework Convention on Climate Change) and desertification (Convention on Desertification).

However, the literature is equivocal about the coordinating and catalytic role played by the UN, pointing out institutional shortcomings and arguing for a stronger anchoring body in global environmental governance \citep{biermann2004assessing, ivanova2010unep, mee2005role}. 

We can test these claims by filtering out UN-sponsored treaties and reconstructing the network without them. The result suggests that the impact of the UN on the network structure has indeed primarily been indirect. UN-sponsored treaties have had little impact on the number of countries in the network (Fig.~\ref{fig:nodes_country}). The majority of countries remain engaged, even with the simulated removal of the UN treaties. The number of statistically significant cooperative links decreases without UN treaties, but not substantially so. 

\subsection{Connectivity}
\begin{figure}[H]
\centering
\begin{subfigure}[c]{0.46\textwidth}
\centering
\includegraphics[width=\textwidth]{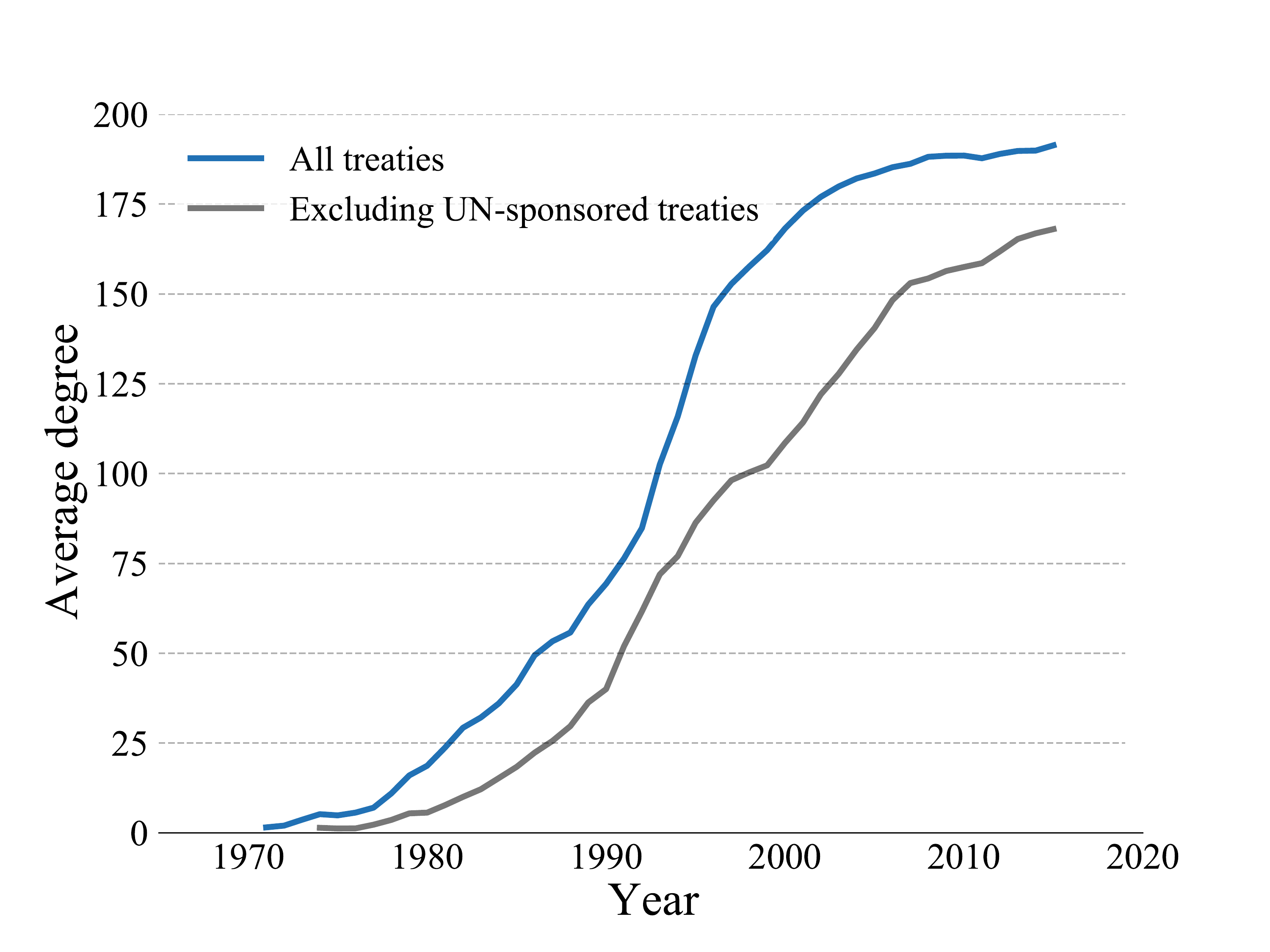}
\caption{Average degree}
\label{fig:average_country_degree}
\end{subfigure}
\begin{subfigure}[c]{0.45\textwidth}
\centering
\includegraphics[width=\textwidth]{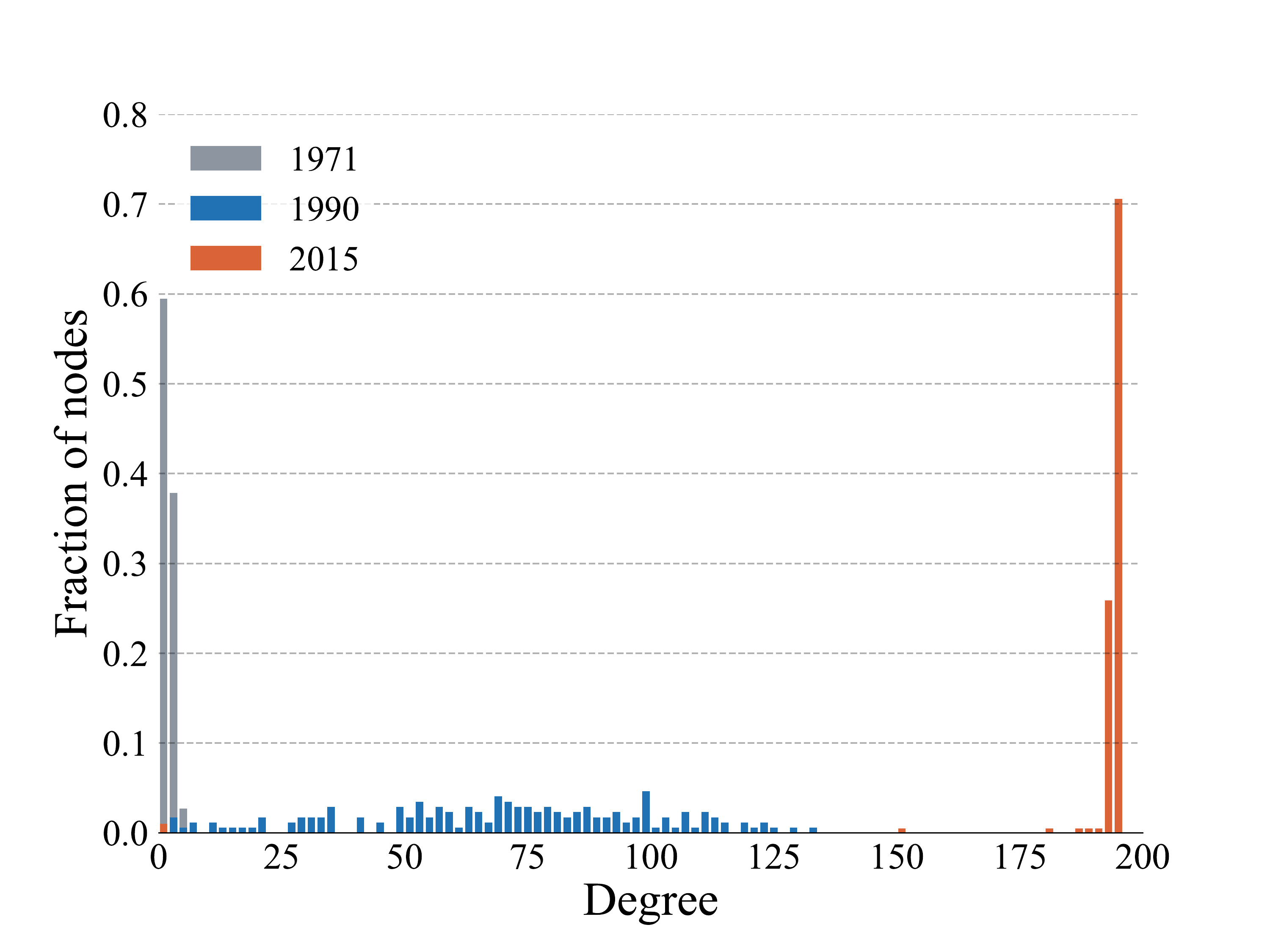}
\caption{Degree distribution}
\label{fig:country_dgeree_frequency}
\end{subfigure}
\vspace*{3mm}
\caption{Average degree and degree distribution from 1971 to 2015.}
\label{fig:country_degree}
\end{figure}

\begin{figure}[H]
\centering
\begin{subfigure}[c]{0.46\textwidth}
\centering
\includegraphics[width=\textwidth]{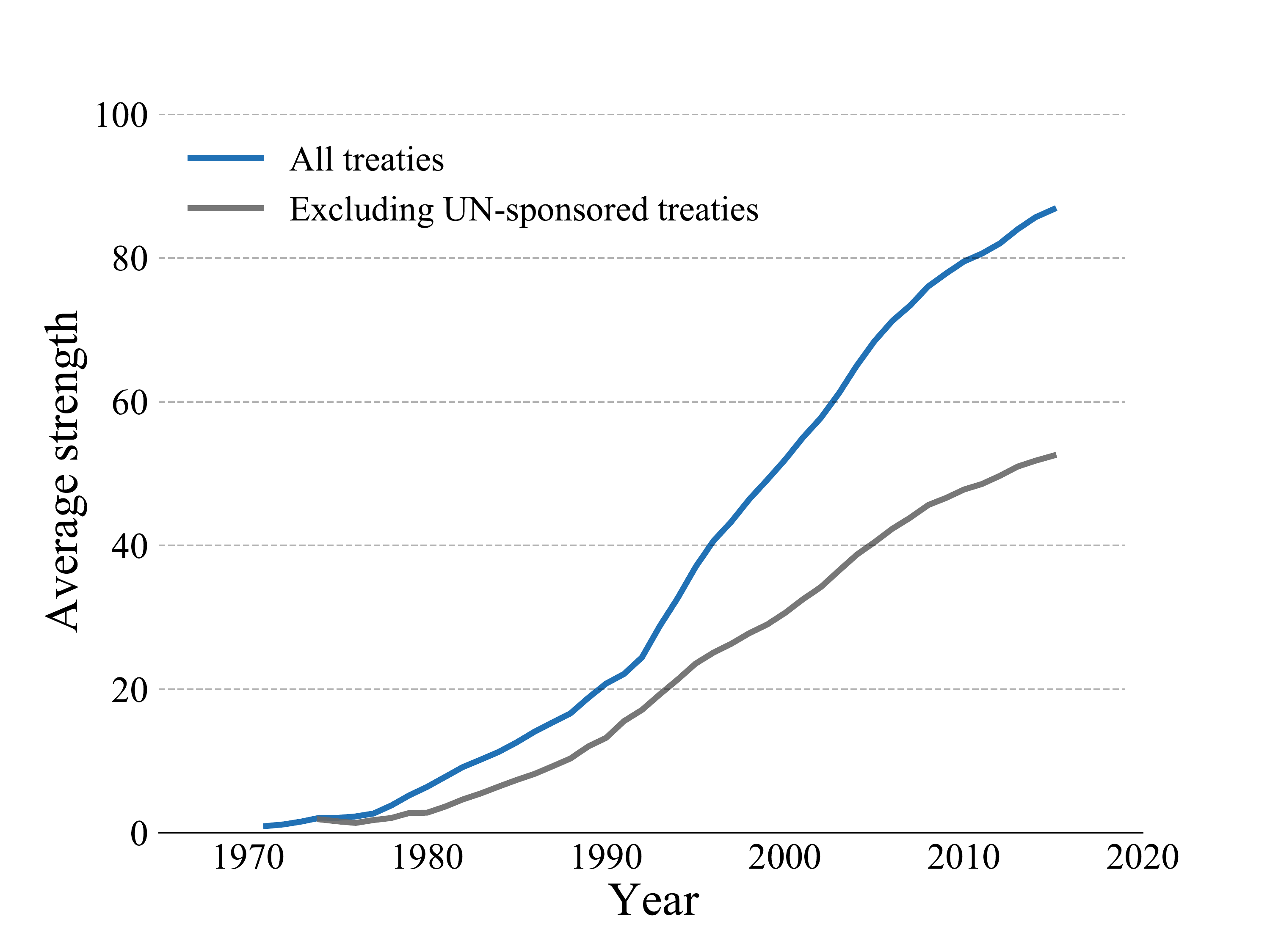}
\caption{Average strength}
\label{fig:average_country_strength}
\end{subfigure}
\begin{subfigure}[c]{0.45\textwidth}
\centering
\includegraphics[width=\textwidth]{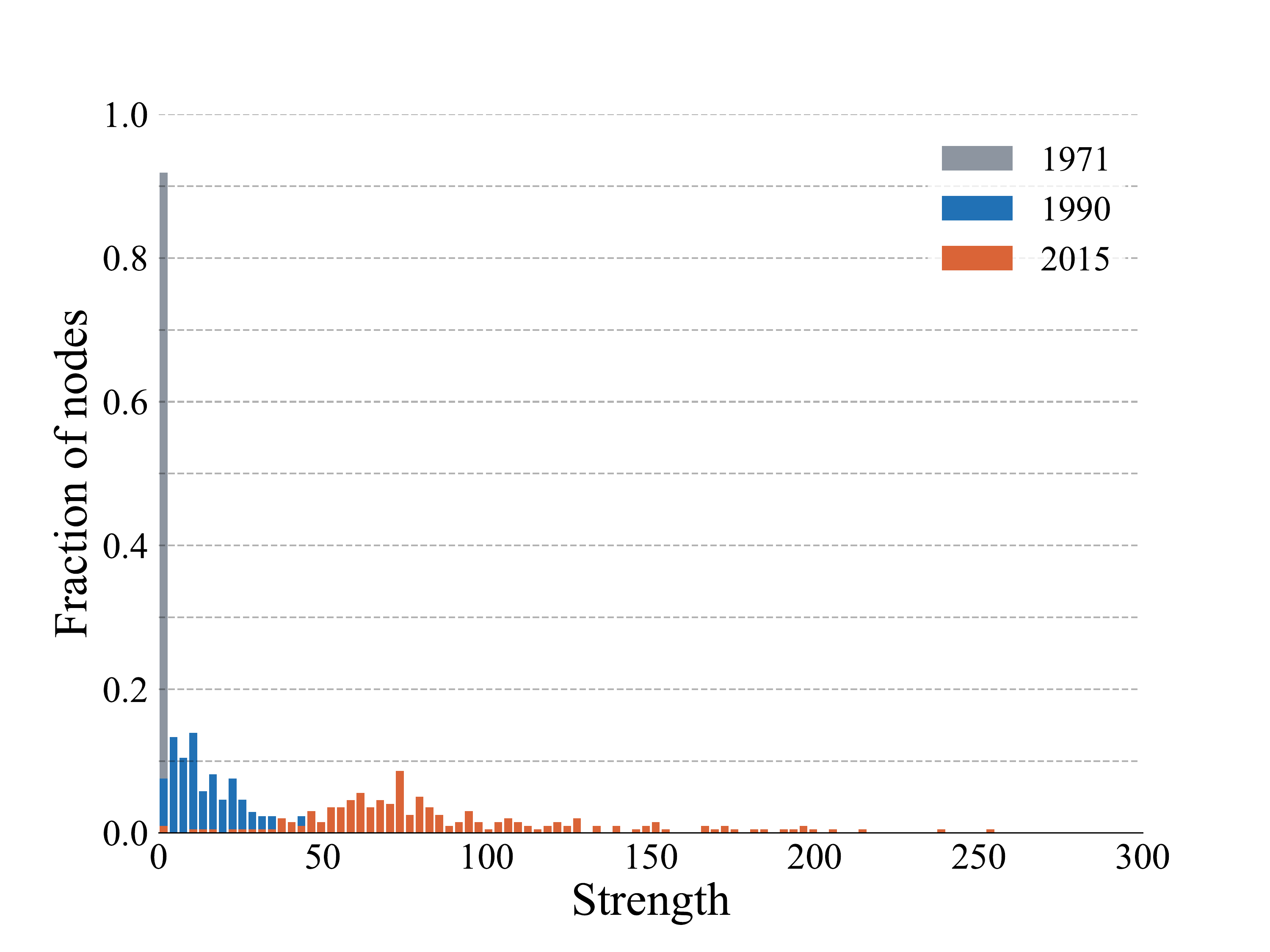}
\caption{Strength distribution}
\label{fig:country_strength_frequency}
\end{subfigure}
\vspace*{3mm}
\caption{Average strength and strength distribution from 1971 to 2015.}
\label{fig:country_strength}
\end{figure}

Over the period of interest, both the average degree and the average strength in the cooperation networks have increased greatly, as shown in panel (a) of Fig.~\ref{fig:country_degree} and ~\ref{fig:country_strength}. The growth in connectivity was particularly pronounced in the 1990s. During this period the degree distribution and strength distribution both widened (panel b), suggesting that the growth in connectivity was initially driven by a vanguard of particularly active countries that forged ahead. By 2015, the degree distribution had narrowed again as the laggards caught up and the average number of partner countries reached a maximum. However, the strength distribution continues to be wide. The cooperation network had reached a point in which connectivity did not depend on the average number of partners, but was constantly reinforced by the average intensity of cooperation among countries.

We again study the impact of UN-sponsored treaties on this pattern by recalculating the metrics for a cooperation network without UN treaties. The average degree of the network decreases only marginally in each year (Fig.~\ref{fig:country_degree}, panel a). However, the exclusion of UN-sponsored treaties reduces the number of common treaties between countries and consequently the average strength in the network. The effect is particularly pronounced in the second half of the study period (Fig.~\ref{fig:country_strength}, panel a).

\section{The ease of collaboration} \label{sec:results_cohesion}

\subsection{Overview}

We next study what the proliferation of IEAs implies for the ability of countries to cooperate and the effectiveness with which knowledge and policy are diffused.

IEAs are both the result of environmental cooperation and a facilitator of such cooperation \citep[e.g.,][]{bernstein2012complex}. The shared objectives and agreed actions from environmental cooperation are frequently codified in an IEA, but these IEAs then create the basis for further cooperation by establishing relationships, providing platforms for engagement and setting up organisational structures to share the benefits of cooperation~\citep[e.g.,][]{meyer1997structuring,bernauer2010comparison,sauquet2014exploring,keohane2005after}. Cooperation through IEAs also creates trust, which is considered key to deal not only with local environmental dilemmas \citep{ostrom_governing_1990}, but also with transnational and global issues \citep{ostrom2009polycentric, owen_trust_2008, carattini_unconventional_2015, hovi_hope_2015, helland_climate_2018,  carattini2019cooperation, carattini_managing_2021}.

The environmental cooperation network also serves as an information network \citep{lazer2005regulatory}, which makes it easier for countries to produce and process information. It has been argued that IEAs, as pieces of international law, are an important driver of policy convergence by helping to disseminate environmental policy knowledge and promote policy diffusion and learning \citep{busch2005global, doi:10.1080/14693062.2014.1000814, holzinger2008environmental}.

We assess these facilitating functions of IEAs by studying the global and local social cohesion of the environmental cooperation network. For the analysis of global cohesion, we will refer to the concepts of components, network density, shortest path length, and global clustering coefficient. To recall, the number of components in a network can be used to gauge the degree of global cohesion across the network, i.e., a network with more components is less cohesive and more fragmented. The network density measures the portion of the potential cooperative connections that are actual connections through treaties. The shortest path length describes the ease and cost of cooperation between countries induced by their structural positions. All else being equal, a network with a small number of components, a high density and a small average shortest path length has a high level of global cohesion and low fragmentation. 

It has been suggested that clustering fosters members' sense of belonging to a shared group~\citep{portes1993embeddedness}, mutual trust, the enforcement of social norms, and the exchange of complex and proprietary information, which in turn may facilitate coordination, cooperation and collective action~\citep{coleman1988social}. Clustering captures social cohesion both at the global and local levels. At the global level, the global clustering coefficient detects the degree to which connected triads tend to close up into triangles across the network. At the local level, the local clustering coefficient captures the tendency of a node's neighbours to become connected themselves. Both measures can be used to uncover closed structures as sources of social capital, and in particular the tendency of collaboration to originate from tightly-knit communities (global level) and third-party relationships (local level). 

Our analysis suggests the following stylised fact.
\\
\textbf{Stylised Fact 2:} \textit{Over the past decades, the network of environmental cooperation has become closer, denser, and more cohesive. Countries have become gradually less isolated when dealing with environmental problems. The network now consists of just one component that connects all countries. The combination of high cohesion at both the global and local levels (high density, short path lengths and high clustering) creates a system that is conducive to policy coordination and the diffusion and exchange of knowledge.} 

\subsection{Cohesion}

In the early 1970s, when statistically significant environmental cooperation links began to emerge, the network consisted of just $37$ countries which formed as many as $12$ components. Practically all of the components were regional groups (for example, there was a component of Middle-Eastern countries) and many were bilateral, consisting of just two nodes with a statistically significant link. The network was small and fragmented. 

By the early 1980s, the cooperation network had grown to $157$ countries which were integrated into a single component. New components formed in the late 1980s and early 1990s as the countries of Eastern Europe and the former Soviet Union started to engage in environmental cooperation. 
For example, in 1991, newly-independent Armenia, Azerbaijan, Georgia, Kazakhstan, Tajikistan, Latvia, and Uzbekistan joined the cooperation network as a separate component. 
They were absorbed into the largest component in the following year, when the network coalesced again into a single global component. Since 1992 every pair of countries (except Taiwan and later Hong Kong) has been able to reach each other through direct or indirect treaty-based connections.

\begin{figure}[H]
\centering
\begin{subfigure}[c]{0.46\textwidth}
\centering
\includegraphics[width=\textwidth]{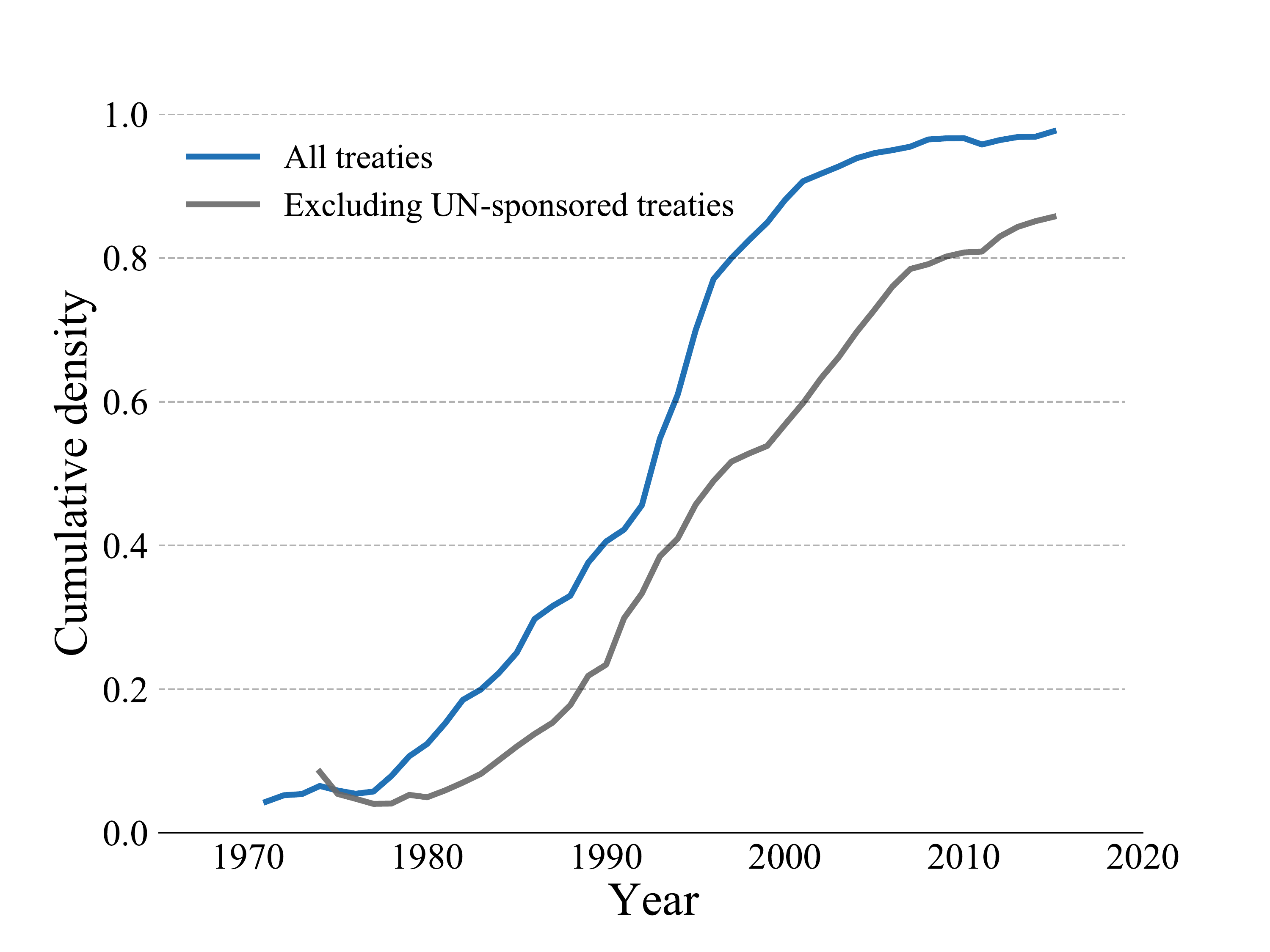}
\caption{Cumulative density}
\label{fig:density_country}
\end{subfigure}
\begin{subfigure}[c]{0.46\textwidth}
\centering
\includegraphics[width=\textwidth]{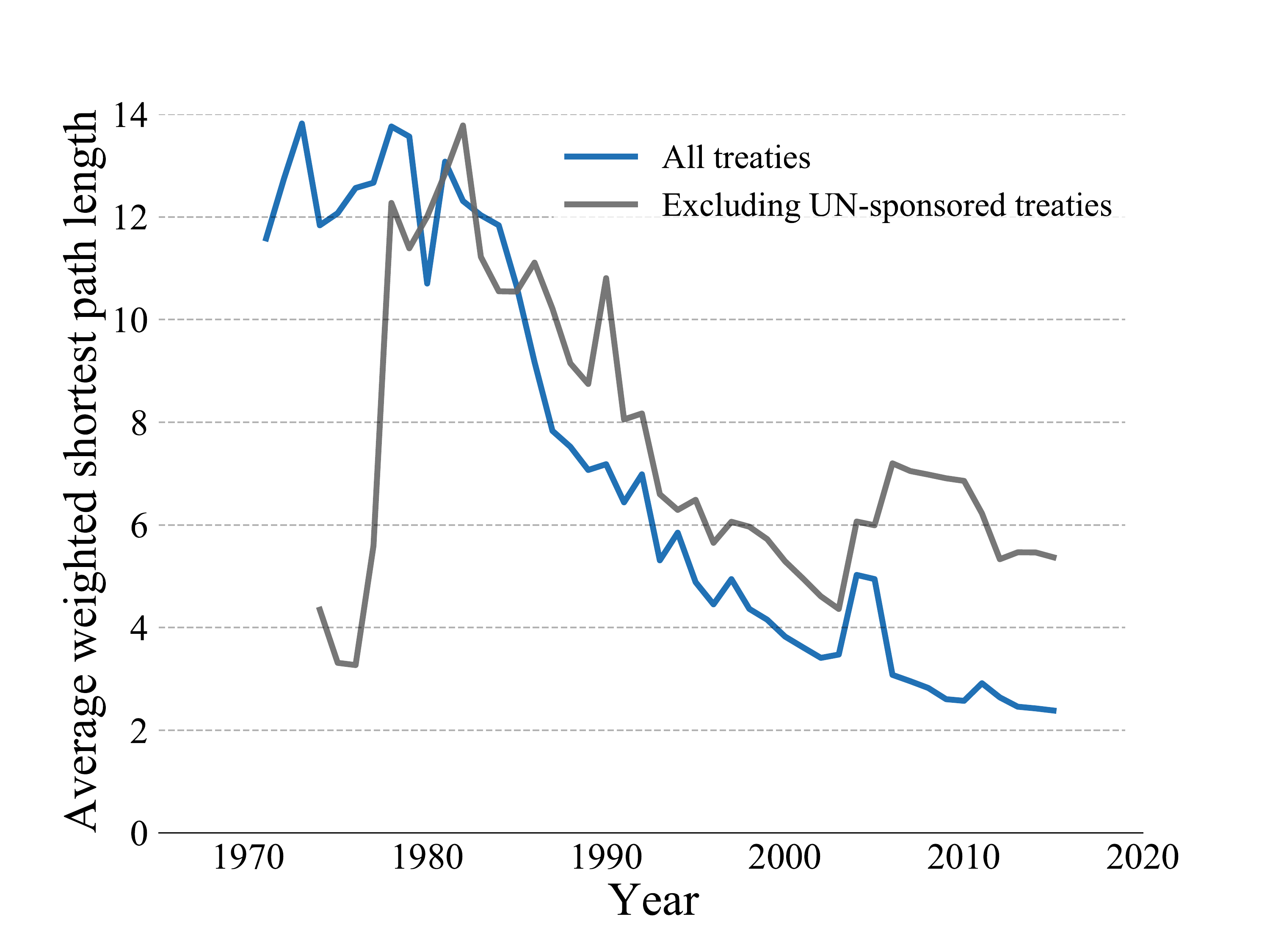}
\caption{Average shortest path length}
\label{fig:path_length_country}
\end{subfigure}
\vspace*{3mm}
\caption{Cumulative density and average shortest path length from 1971 to 2015.}
\label{fig:density_path_country}
\end{figure}

The density of the cooperation network grew at a similar pace, increasing rapidly through the 80s and 90s. The network began to stabilise at the start of this century, at which point nearly every pair of countries had established a significant cooperation relationship (Fig.~\ref{fig:density_path_country}, panel a). 

The average shortest path length stayed at a high level in the 1970s of the cooperation network, reflecting the growing size of the dominant network component, but has fallen steadily since (Fig.~\ref{fig:density_path_country}, panel b). The size of the dominant network component remained stable throughout this period, encompassing some $95$ percent of nodes. At the same time new connections appeared and existing connections were strengthened through new treaties, which in turn fostered a reduction in average path lengths. 
Overall, these results corroborate the view that the fall of the Soviet Union and the end of the Cold War in the early 1990s created the opportunity for new alliances, encouraging international cooperation and policy diffusion to occur outside the two hegemonic blocks~\citep{Yamagata}. 

Although the exclusion of UN-sponsored treaties leads to a lower density, as shown in panel (a) of Fig.~\ref{fig:density_path_country}, the overall trend does not change without the UN treaties. The situation is the same for the average weighted shortest path length after 1980.
Before 1980, the exclusion of UN treaties leads to more components and a smaller fraction of the largest component (less than $30\%$), which results in lower average shortest path length. 
However, even without the UN treaties the whole network remains connected from 1990 to 2015. Thus, while the UN has contributed to reducing the distance between countries and, as a result, the cost of cooperation among them, we conclude that cooperation outside UN platforms has been as dense and effective as under the aegis of the UN. This result talks to the debate in the literature and reconciles the two opposing views on the role played by the UN; on the one hand we find that UN contributed to the creation of a world environmental regime, by providing an organisational context and framework to promote inter-state cooperation and with an agenda broad enough to include environmental issues~\citep{meyer1997structuring}. On the other hand, the cooperation network emerged and developed also outside the forums provided by the UN.

\subsection{Clustering}

\begin{figure}[H]
\centering
\includegraphics[scale=0.3]{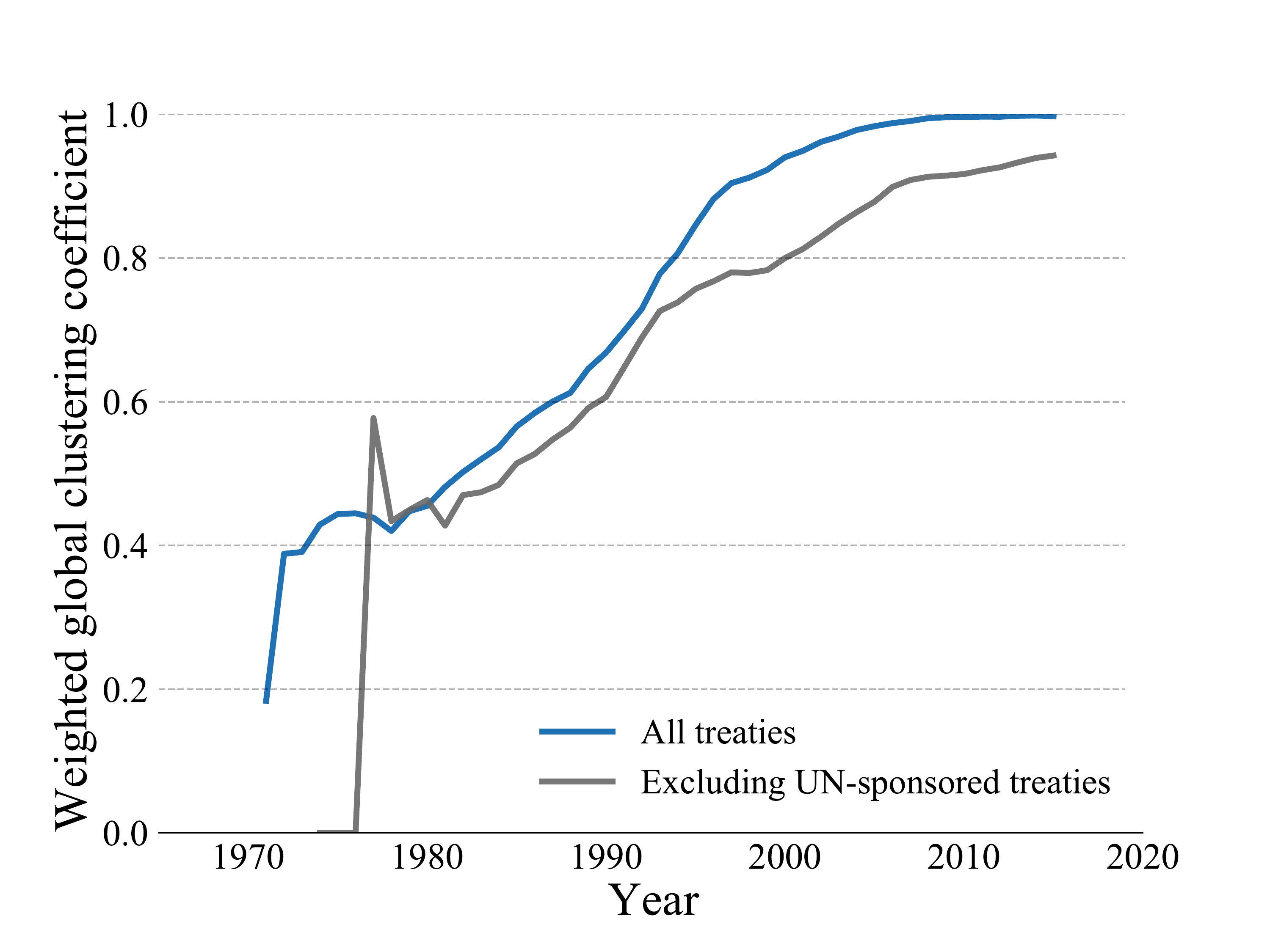}
\caption{Global clustering coefficient from 1971 to 2015.}
\label{fig:clustering_country}
\end{figure}

The evolution of the global clustering coefficient of the network is shown in Fig.~\ref{fig:clustering_country}. Following a short blip in the 1970s, the clustering coefficient has grown rapidly and steadily through the 1980s and 1990s before levelling off at the beginning of this century. As such, the trend is comparable to that observed for the network size and connectivity metrics. It suggests that, as the cooperation network expanded and new links were created, third-party relationships (i.e., links between countries sharing partners) were formed simultaneously and at the same rate.

Many factors can promote the presence of common partners, such as geographic proximity, affiliation with related regional groups or organisations, a similar economic status, a shared history and trading relationships~\citep{fagiolo2010evolution,sauquet2014exploring}. The presence of common partners is likely to have promoted trust and helped countries to establish deeper relationships.

As we have observed with other network metrics, the overall trend of the global clustering coefficient does not change significantly when UN-sponsored treaties are removed. 

%Not sure if we still need this figure or what it adds?

%\begin{figure}[H]
%\centering
%\begin{subfigure}[c]{0.45\textwidth}
%\centering
%\includegraphics[width=\textwidth]{images/country/CountryMap_1970_d.pdf}
%\caption{1970}
%\label{}
%\end{subfigure}
%\begin{subfigure}[c]{0.45\textwidth}
%\centering
%\includegraphics[width=\textwidth]{images/country/CountryMap_1980_d.pdf}
%\caption{1980}
%\label{}
%\end{subfigure} \\
%\begin{subfigure}[c]{0.45\textwidth}
%\centering
%\includegraphics[width=\textwidth]{images/country/CountryMap_1990_d.pdf}
%\caption{1990}
%\label{}
%\end{subfigure}
%\vspace*{-1.2mm}
%\begin{subfigure}[c]{0.45\textwidth}
%\centering
%\includegraphics[width=\textwidth]{images/country/CountryMap_2000_d.pdf}
%\caption{2000}
%\label{}
%\end{subfigure} \\
%\begin{subfigure}[c]{0.45\textwidth}
%\centering
%\includegraphics[width=\textwidth]{images/country/CountryMap_2010_d.pdf}
%\caption{2010}
%\label{}
%\end{subfigure}
%\begin{subfigure}[c]{0.45\textwidth}
%\centering
%\includegraphics[width=\textwidth]{images/country/CountryMap_2015_d.pdf}
%\caption{2015}
%\label{}
%\end{subfigure}
%\caption{\textcolor{brown}{Country networks in 1970, 1980, 1990, 2000, 2010 and 2015. The size of a node is proportional to its strength, and the colour of a node is proportional to its local clustering coefficient. Except the map in 1970, other maps only show top $10$ percent of edges in terms of weights for better visualisation.}}
%\label{fig:country_maps}
%\end{figure}

\section{The strategic role of countries}\label{sec:results_by_countries}

\subsection{Overview}

We now turn to the positions of individual countries in the cooperation network. The roles of different countries in environmental cooperation is an important subject in the international relations literature,  covering angles such as the influence of hegemons \citep{Yamagata} and the changing role of players like the United States~\citep{falkner2005american,kelemen2010trading} and the European Union ~\citep{falkner2007political,kelemen2010globalizing, vogler2007european}. 

To measure the role of individual countries we use the centrality metrics of node strength, betweenness centrality and closeness centrality. To recall, node strength accounts for the intensity of cooperation of a country with others, while betweenness centrality measures the ability of a country to intermediate between others. Closeness centrality measures the distance of a focal country to the other countries in the network, and consequently the potential cost for further cooperation, based on existing treaty-based connections.

We do not report node degrees. Although the node degree of a country is an important centrality measure, which quantifies the number of partners with which a country cooperates, the metric is less important in our case. Because of the high density of the environmental cooperation network, there is little heterogeneity in node degrees. The weight of links needs to be taken into account, which makes node strength the better metric to quantify node heterogeneity. The issue is exacerbated by the presence in the network of a number of large treaties with almost global participation, which leads to a uniformly high degree for signatory countries. The phenomenon is particularly pronounced in the case of air and atmosphere treaties (see Section~\ref{sec-results_by_treaties} below).

The analysis gives rise to \textbf{Stylised Fact 3:} \textit{The network of environmental cooperation is fundamentally global, but with a noticeable European imprint. In terms of cooperation intensity, closeness and brokerage power, the network is heavily controlled by European countries, in particular the United Kingdom and more recently France and Germany. The strategic position of countries has remained relatively stable over time, although there are important fluctuations. They include the gradual decline in the network centrality of Japan and the US, and the emergence of South Africa as a strategic node among African countries. } 

\subsection{Centrality}

We find strong path dependence in the strategic role of individual countries in the cooperation network. The countries that topped the centrality rankings at the outset were broadly able to  maintain their strategic positions. This stability is in contrast to other networks, where the  centrality of individual nodes is often highly sensitive to changes in the network structure (in our case, the signing of new treaties).

We assess the stability of countries' network position over time by looking at the Kendall-Tau correlation coefficients of country rankings for different centrality measures. Kendall-Tau measures the rank correlation for each centrality measure between time window $t$ and $t+1$.  The starting point of the analysis is the year 1980, when the number of countries in the network begins to stabilise (see Fig.~\ref{fig:nodes_country} above) and the rankings of countries are comparable. 

For each centrality measure we find a statistically significant and positive correlation between country rankings over time. The path dependence is most pronounced in the case of strength and closeness centrality, with Kendall Tau coefficients of around $0.9$. The positive correlation for betweenness centrality is lower, but has solidified over time, from $0.65$ to $0.85$. 

%\begin{figure}[H]
%\centering
%\includegraphics[scale=0.3]{country/Kendall_coef_2.pdf}
%\caption{Kendall'tau correlation coefficient between time window $t$ and $t+1$ for degree, strength, betweenness centrality and closeness centrality. The time window is two years, and in each time window each centrality measure of a country is averaged over two years. The results are statistically significant.}
%\label{fig:kendall}
%\end{figure}

\begin{figure}[H]
\centering
\begin{subfigure}[c]{0.9\textwidth}
\centering
\includegraphics[width=\textwidth]{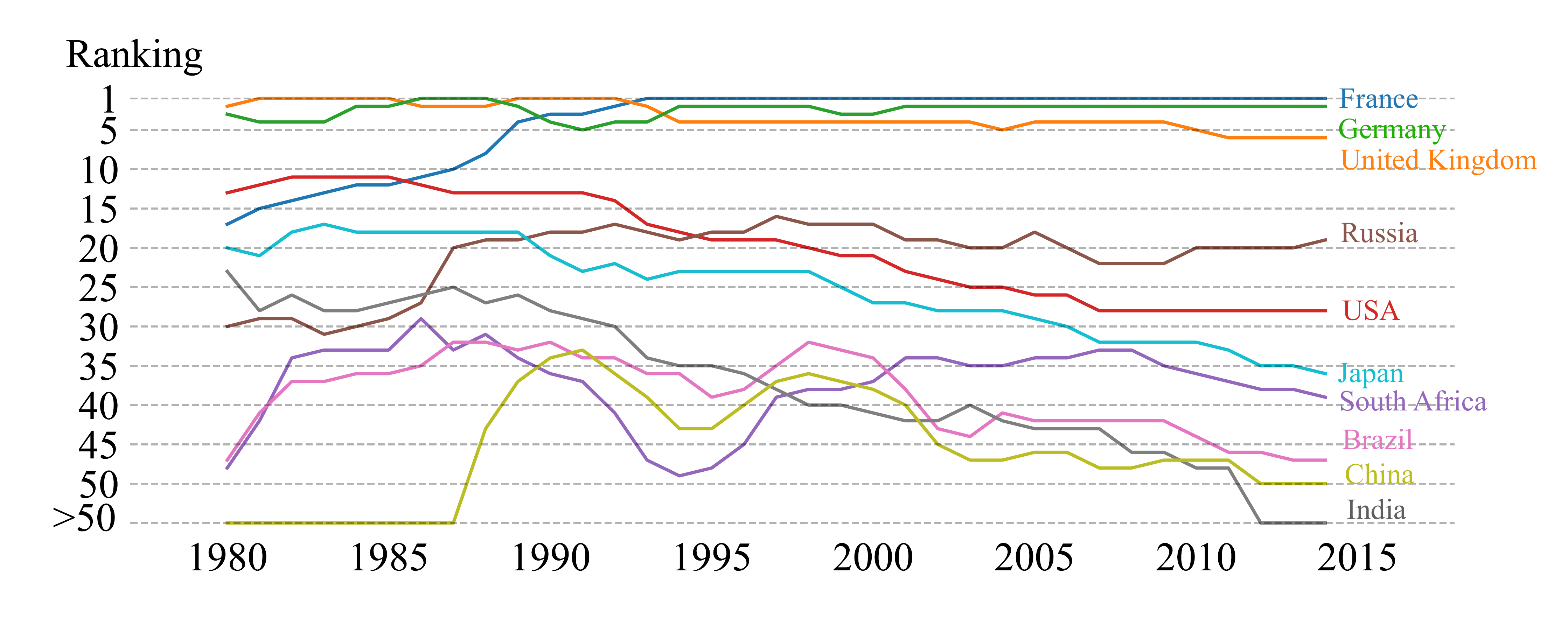}
\vspace{-10mm}
\caption{Strength of countries}
\label{fig:strengthranking_country}
\end{subfigure}
\begin{subfigure}[c]{0.9\textwidth}
\centering
\includegraphics[width=\textwidth]{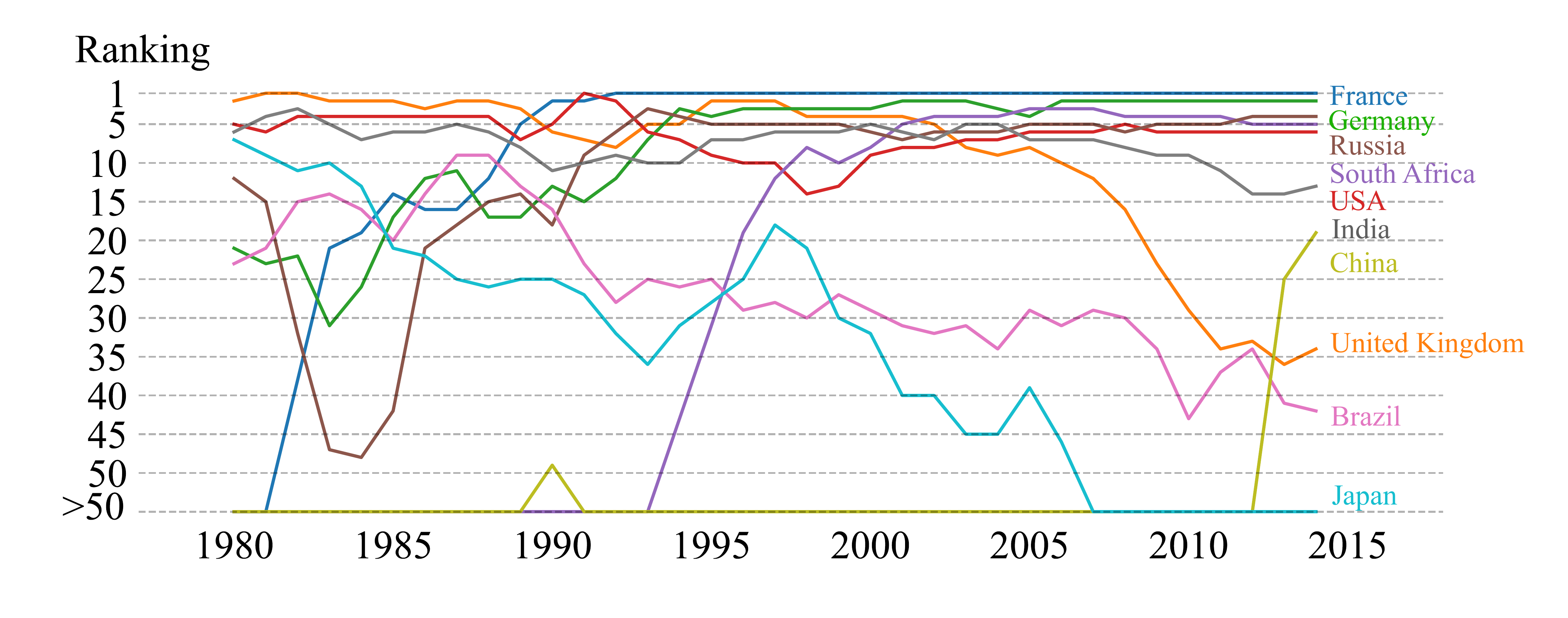}
\vspace{-10mm}
\caption{Betweenness centrality of countries}
\label{fig:betweennessranking_country}
\end{subfigure}
\begin{subfigure}[c]{0.9\textwidth}
\centering
\includegraphics[width=\textwidth]{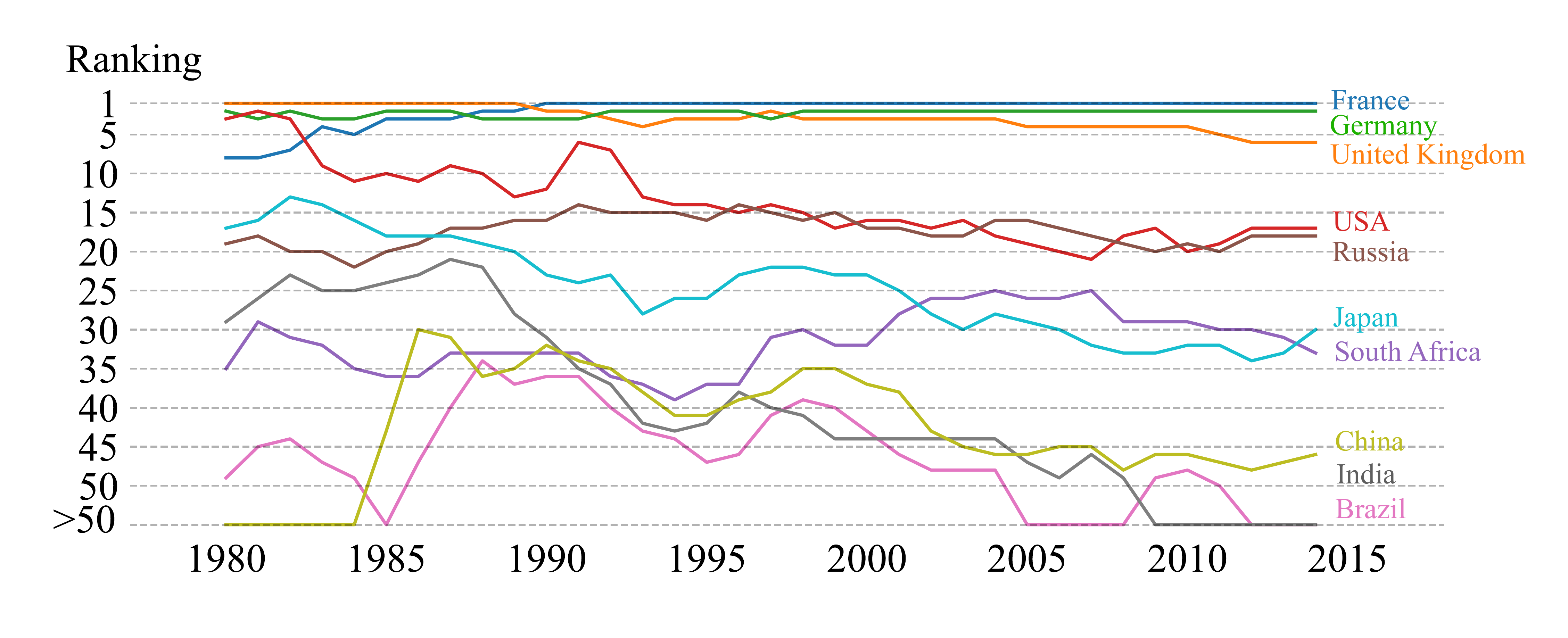}
\vspace{-10mm}
\caption{Closeness centrality of countries}
\label{fig:closenessranking_country}
\end{subfigure}
\caption{Centrality measures. Country rankings from 1980 to 2015.}
\label{fig:centralityranking_country}
\end{figure}

Within this stable overall pattern it is possible to discern some notable trends for individual countries. While our methodology accentuates smaller countries, we are interested in particular in the network position of major economies. Fig.~\ref{fig:centralityranking_country} shows overall trend in our chosen metrics for 10 major economies: five members of the G7 (Germany, France, United Kingdom, Japan and the USA), the four BASIC countries (Brazil, China, India and South Africa) and Russia. The statistics are shown in terms of country rankings, since we are interested in the relative position of countries, rather than the actual centrality scores. 

The strongest positions in the network are held by European countries, which have both high node strengths and centrality scores. For the past few years, France and Germany were  ranked first and second with respect to all three centrality measures. This makes the two countries significant hubs in environmental cooperation, with a high cooperation intensity, significant brokerage power and, thanks to the short network distance to other countries, the ability to exert control over the circulation of information in the cooperation network. 

France and Germany are replacing the United Kingdom at the top of the rankings. The United Kingdom played a dominant network role in the 1980s and continues to be a hub in terms of cooperation intensity (node strength). However, its position as a network broker (betweenness centrality) is waning.

Reflecting its recent ambivalence to international environmental cooperation, the network centrality of the US has decreased notably over the years. The US still exerts considerable influence over the network, but does not play the dominant role one might expect from a global super power. The final G7 country, Japan, has also seen its influence wane. 

These rankings corroborate a widely held view in the international relations literature, which speaks of a shift in international environmental leadership from the US to the EU \citep{vogler2007european, kelemen2010trading,kelemen2010globalizing}. The rankings also speak to future prospects. The roles of different countries in the cooperation network are both a reflection of their past behaviours in international environmental politics and an indicator of future strengths or weaknesses when seeking international cooperation.

% Old table with the actual values. JianJian and Sam prefer to only include the rankings here, as it is the rankings of countries that matters, not the actual values. }
%\begin{figure}[H]
%\centering
%\begin{subfigure}[c]{0.46\textwidth}
%\centering
%\includegraphics[width=\textwidth]{country/Degree_country.pdf}
%\caption{Degree of countries}
%\label{fig:degree_per_country}
%\end{subfigure}
%\begin{subfigure}[c]{0.46\textwidth}
%\centering
%\includegraphics[width=\textwidth]{country/Strength_country.pdf}
%\caption{Strength of countries}
%\label{fig:strength_per_country}
%\end{subfigure}
%\begin{subfigure}[c]{0.46\textwidth}
%\centering
%\includegraphics[width=\textwidth]{country/Betweenness_country.pdf}
%\caption{Betweenness of countries}
%\label{fig:betweenness_per_country}
%\end{subfigure}
%\begin{subfigure}[c]{0.46\textwidth}
%\centering
%\includegraphics[width=\textwidth]{country/Closeness_country.pdf}
%\caption{Closeness of countries}
%\label{fig:closeness_per_country}
%\end{subfigure}
%\vspace*{3mm}
%\caption{Centrality measures from 1968 to 2015.}
%\label{fig:centrality_country}
%\end{figure}

Among emerging markets it is worth noting the growing network power of South Africa. The country has a middling node strength, but is emerging as an increasingly important broker at the centre of the network. Analysis of an Africa-only network suggests that South Africa's role is underpinned by its centrality, post-Apartheid, in African environmental cooperation. 

South Africa's role is in contrast to the low centrality of other emerging markets to the cooperation network, including perhaps most notably China's. Until relatively recently environmental issues were not high on the agenda of the Chinese government, either domestically or internationally, although this is starting to change, for example with an increased domestic interest in  air quality and a stronger international role on climate change ~\citep {green2015china}.

\section{Differences across environmental issues}\label{sec-results_by_treaties}

\subsection{Overview}

Different environmental problems have attracted international attention at different times and with varying intensity. This reflects differences in the interplay between interests, political power and discourse within and between countries ~\citep{mitchell2020we, mitchell2003international}, as well as the distinct characteristics of different environmental problems~\citep{falkner2013handbook}. This makes it important to study the cooperation patterns among countries under different treaty subjects. 

To do so, we construct separate cooperation networks for the different categories of treaties introduced in section \ref{Sec:motivation}. We use the same metrics as in previous sections, with a focus on network size (number of nodes), connectivity (degree, strength), and cohesion (density, shortest path length, clustering coefficient). 

This allows us to describe in topological terms the 'regime complexity' discussed in the international environmental governance literature \citep{meyer1997structuring, keohane_regime_2011}. 

The analysis gives rise to \textbf{Stylised Fact 4:} \textit{Environmental cooperation has distinctly different network features depending on the subject area. Environmental coordination started with the management of marine resources (fisheries and the sea), but is now strongest in the area of waste and hazardous substances. The networks on species, waste and natural resources have a hierarchical structure, where a series of densely connected, small clusters combine into a less dense global network. This feature is absent in the networks on sea and fisheries and air and atmosphere. Despite the high policy salience of the topic, cooperation in the air and atmosphere network appears to be less intensive and the network is less cohesive.  Unlike the other networks, the air and atmosphere network is heavily shaped by UN-sponsored treaties.} 

\subsection{Network properties by treaty subject}

The topic-specific cooperation networks obtained statistical significance at different times. A significant cooperation network first appeared in sea and fishery affairs in 1985, followed by natural resources in 1987, waste and hazardous substances in 1990, wild species and ecosystems in 1994 and air and atmosphere in 2000. Based on our method, the cooperation network for energy treaties does not reach statistical significance, and we therefore do not analyse this network.

The topic-specific networks become statistically significant later than the overall network for methodological reasons. When treaties are divided into different categories, each category has a smaller number of treaties, relative to the number of countries. In some of the early country-treaty bipartite networks, the number of countries can be more than four times the number of treaties. When projecting onto the country layer to obtain the cooperation network, this makes it harder for the number of co-signed treaties between countries to be significantly different from the null model. Our interest is therefore in the sequence in which topic networks become significant and not the specific dates.

The different speed at which international cooperation occurred may reflect a number of factors, including the changing salience of different environmental matters over time (e.g., the emergence of climate change as an issue in the 1990s), path dependency (the deepening of links in areas of long-standing cooperation) and potentially an initial focus on subjects where cooperation is easier \citep[per][] {keohane_regime_2011}. 

However, by 2005 most countries had joined all five cooperation networks, suggesting that countries are now collaborating across the full range of environmental issues. In each subject area, nearly all the countries now form a single component. 

\begin{figure}[H]
\centering
\begin{subfigure}[c]{0.3\textwidth}
\centering
\includegraphics[width=\textwidth]{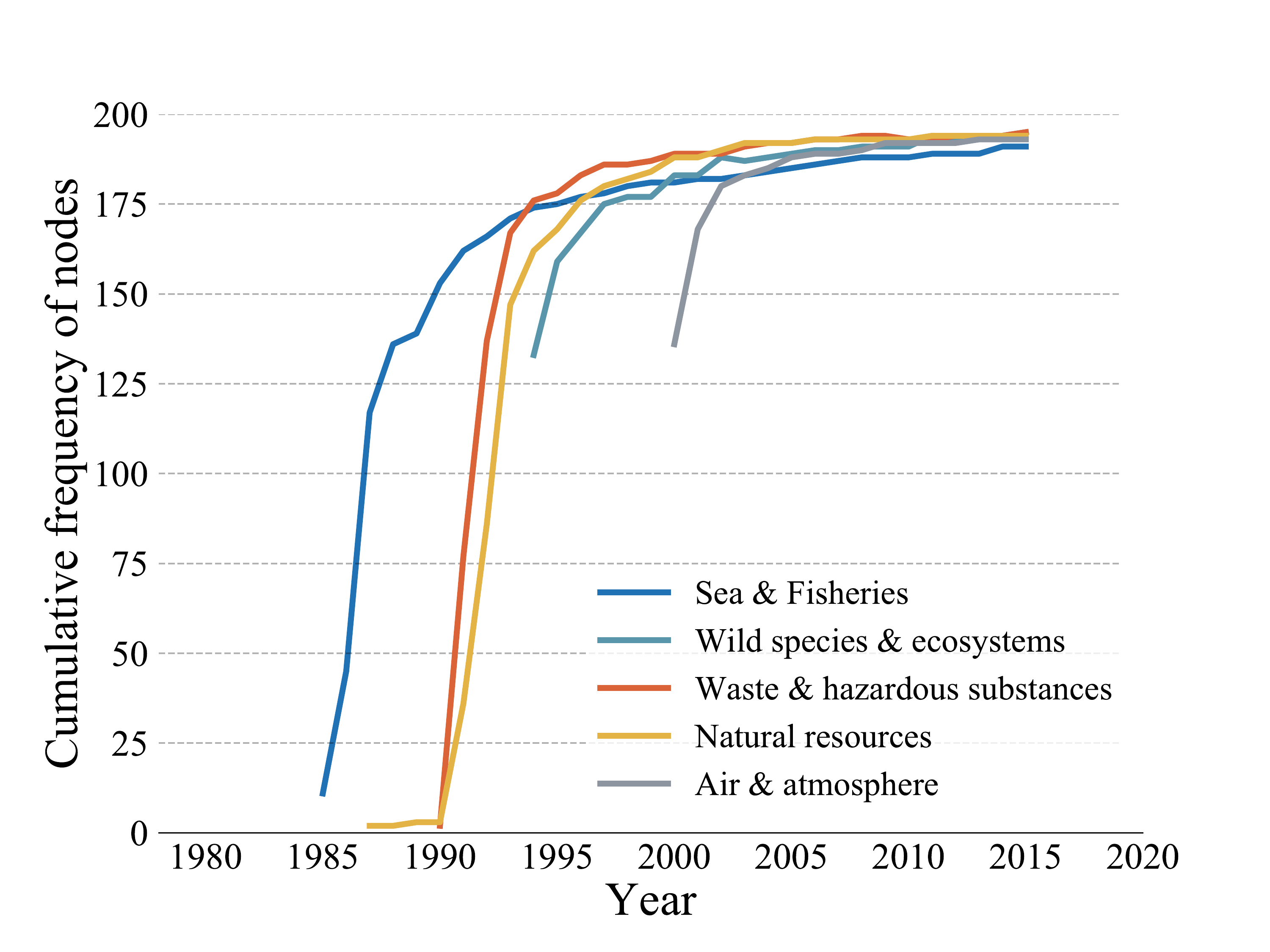}
\caption{Number of nodes}
\label{}
\end{subfigure}
\begin{subfigure}[c]{0.3\textwidth}
\centering
\includegraphics[width=\textwidth]{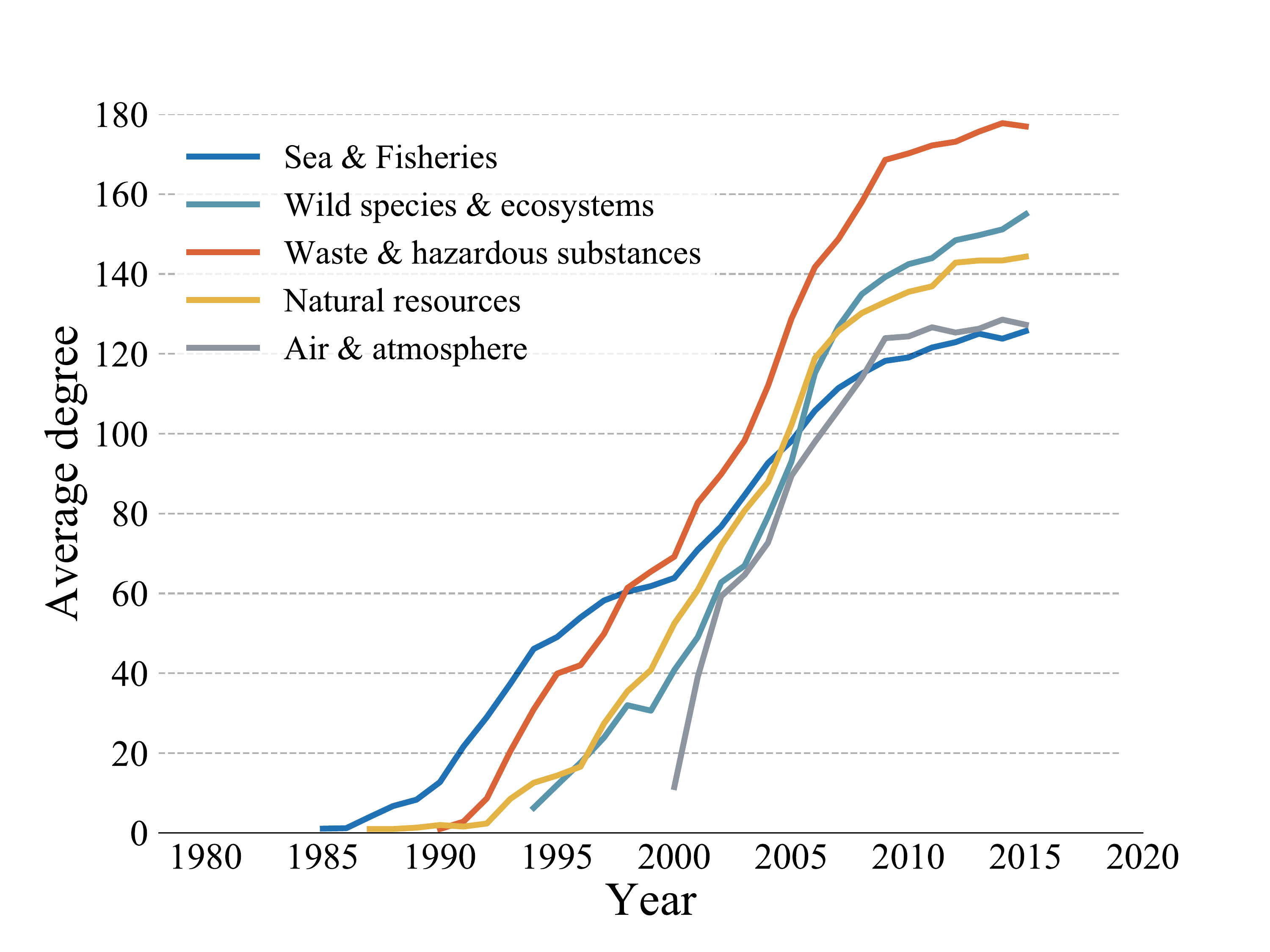}
\caption{Average degree}
\label{fig:degree_country_subjects_test}
\end{subfigure}
\begin{subfigure}[c]{0.3\textwidth}
\centering
\includegraphics[width=\textwidth]{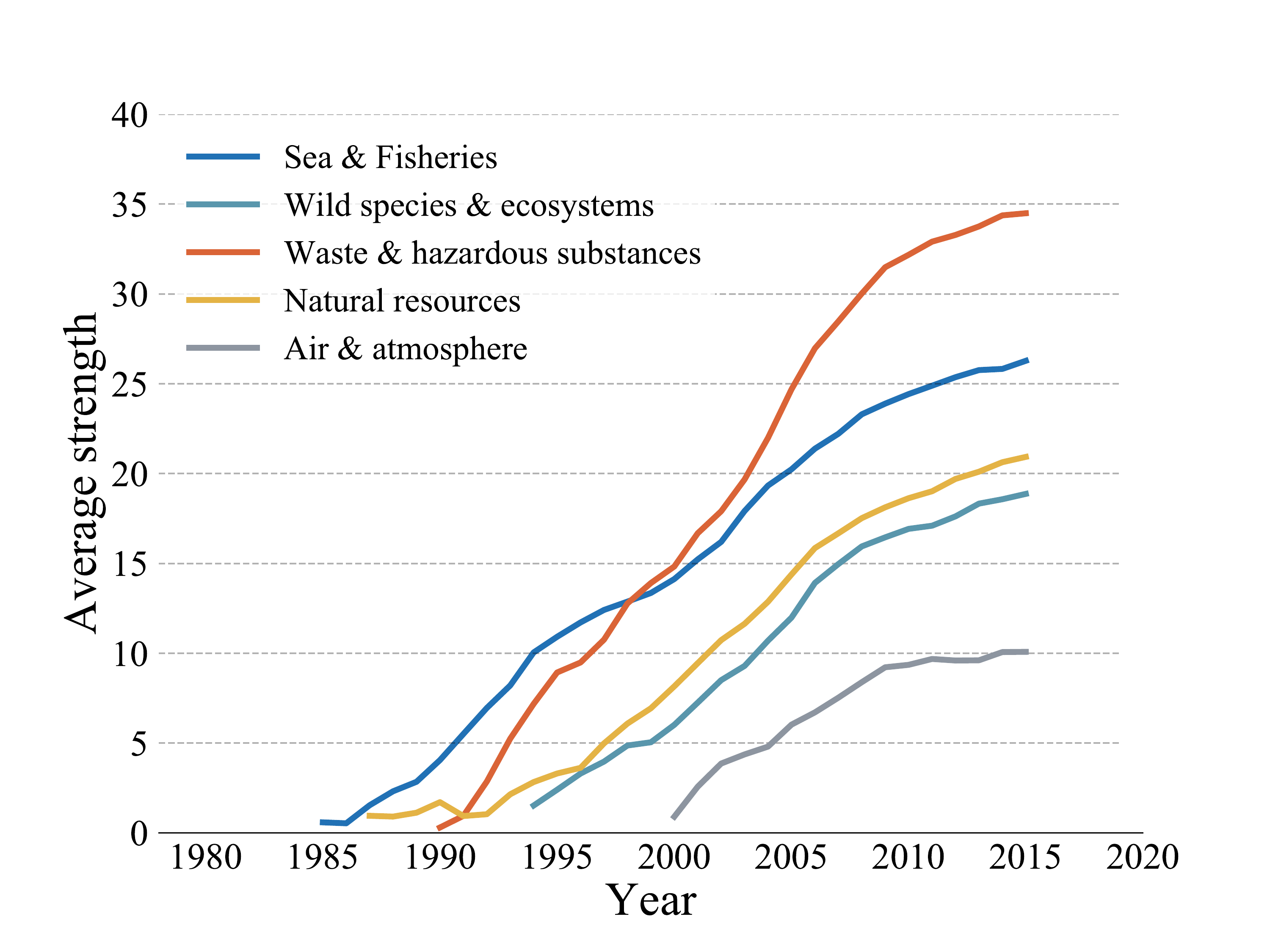}
\caption{Average strength}
\label{fig:strength_country_subjects_test}
\end{subfigure} \\
\begin{subfigure}[c]{0.3\textwidth}
\centering
\includegraphics[width=\textwidth]{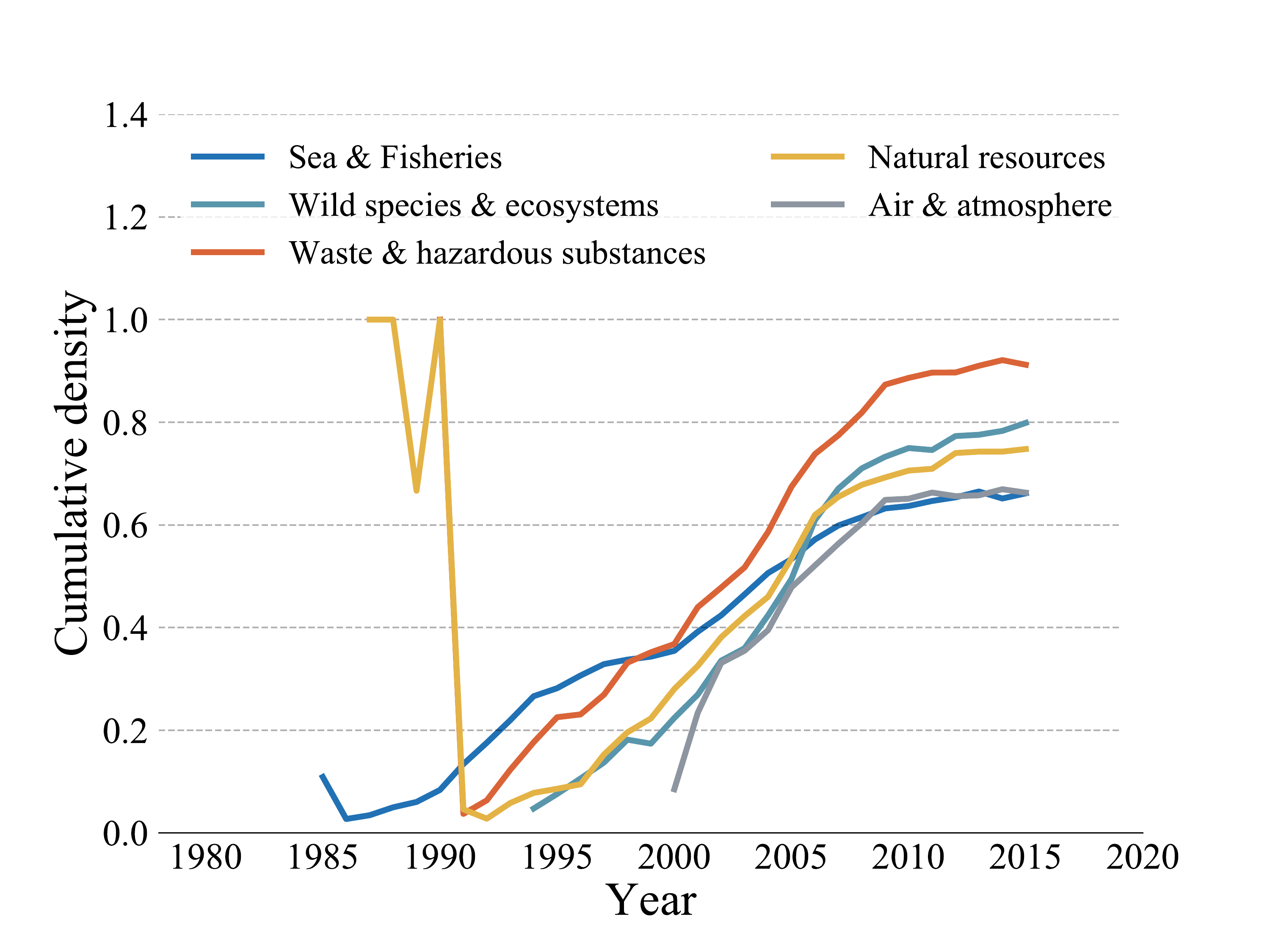}
\caption{Density}
\label{fig:density_subjects}
\end{subfigure}
\begin{subfigure}[c]{0.3\textwidth}
\centering
\includegraphics[width=\textwidth]{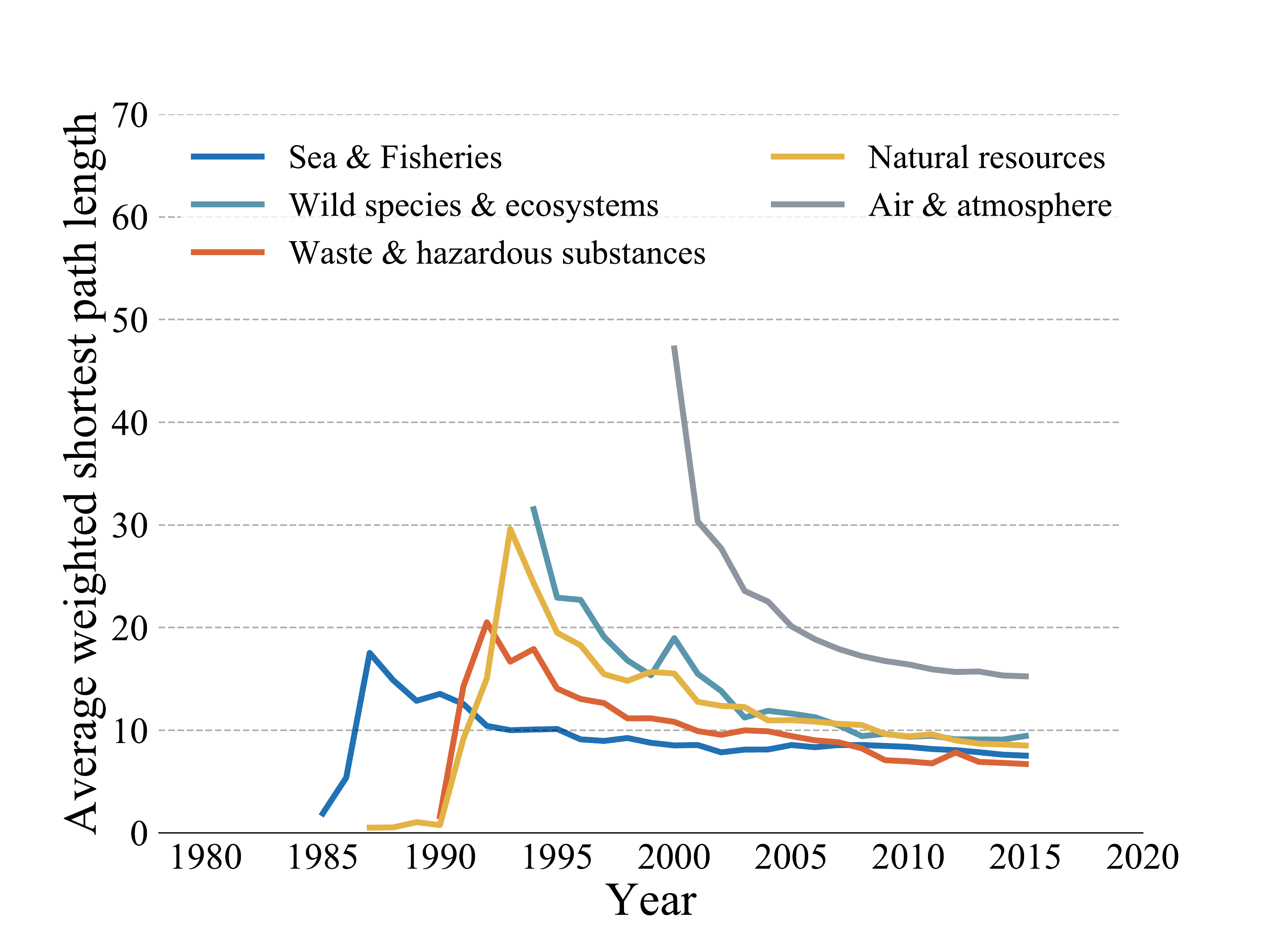}
\caption{Average shortest path length}
\label{fig:path_subjects}
\end{subfigure}
\begin{subfigure}[c]{0.3\textwidth}
\centering
\includegraphics[width=\textwidth]{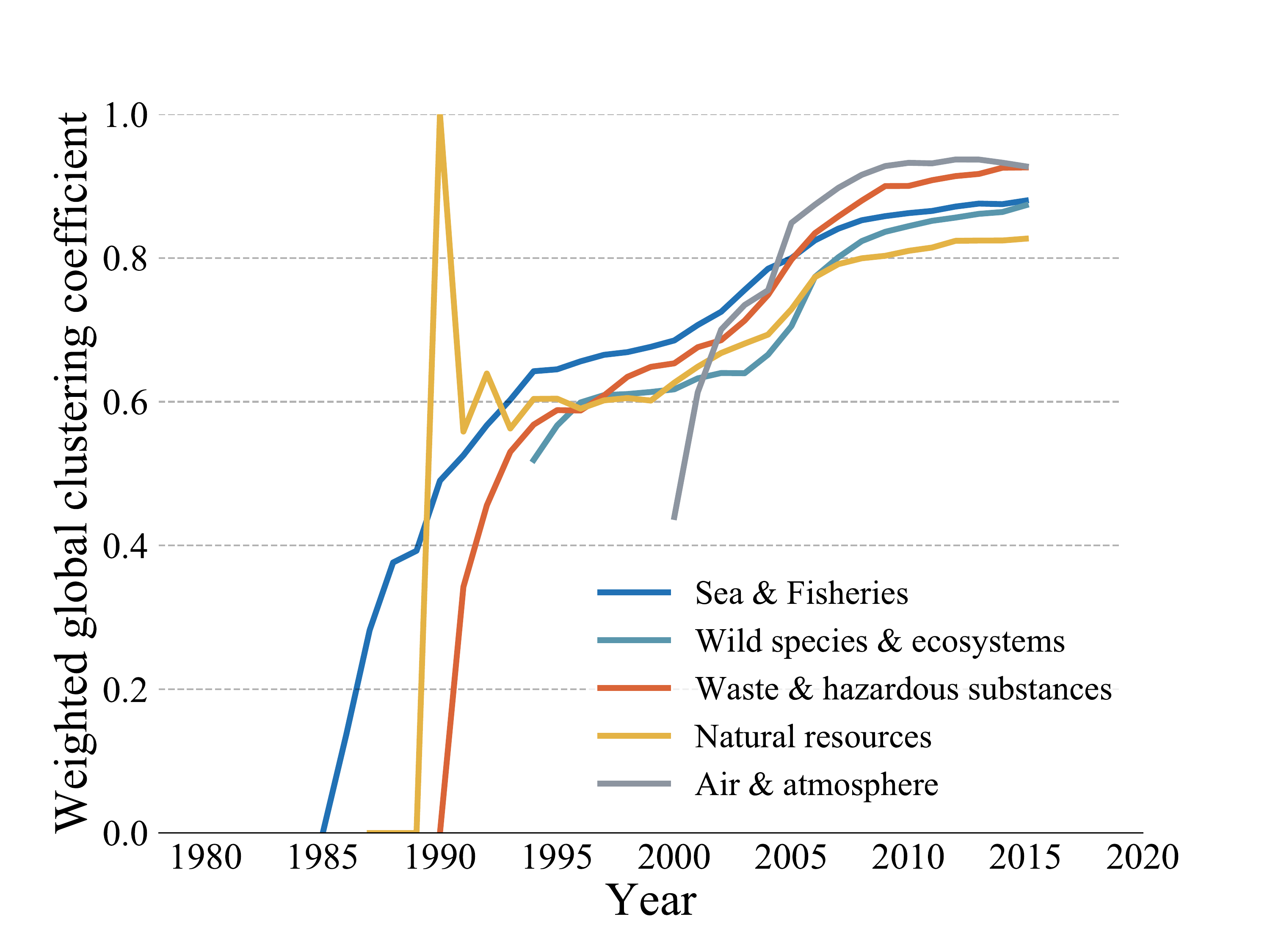}
\caption{Global clustering coefficient}
\label{fig:clustering_subjects}
\end{subfigure}
\vspace*{3mm}
\caption{Cooperation networks for different treaty subjects.}
\label{fig:country_subjects}
\end{figure}

The relative growth in network size and connectivity is shown in Fig.~\ref{fig:country_subjects}. According to the figure, the cooperation network on waste and hazardous substances ranks first in terms of size (number of nodes), connectivity (average degree, average strength), and global cohesion (density, average shortest path length and global clustering coefficient). 

The cooperation network on air and atmosphere is worth a closer look.  Although countries have a high average number of partners in this network, the average cooperation intensity is relatively low. The air and atmosphere network is also globally less cohesive with a low density and a high average shortest path length. In contrast, the network has a high global clustering coefficient, indicating high local density among partners. There are a number of high-profile, often successful treaties with near global membership such as the 1985 Vienna Convention and the 1987 Montreal Protocol \citep{parson2003protecting, falkner2010international}. However, compared with other categories the overall number of air and atmosphere treaties is relatively small.

\begin{figure}[H]
\centering
\begin{subfigure}[c]{0.46\textwidth}
\centering
\includegraphics[width=\textwidth]{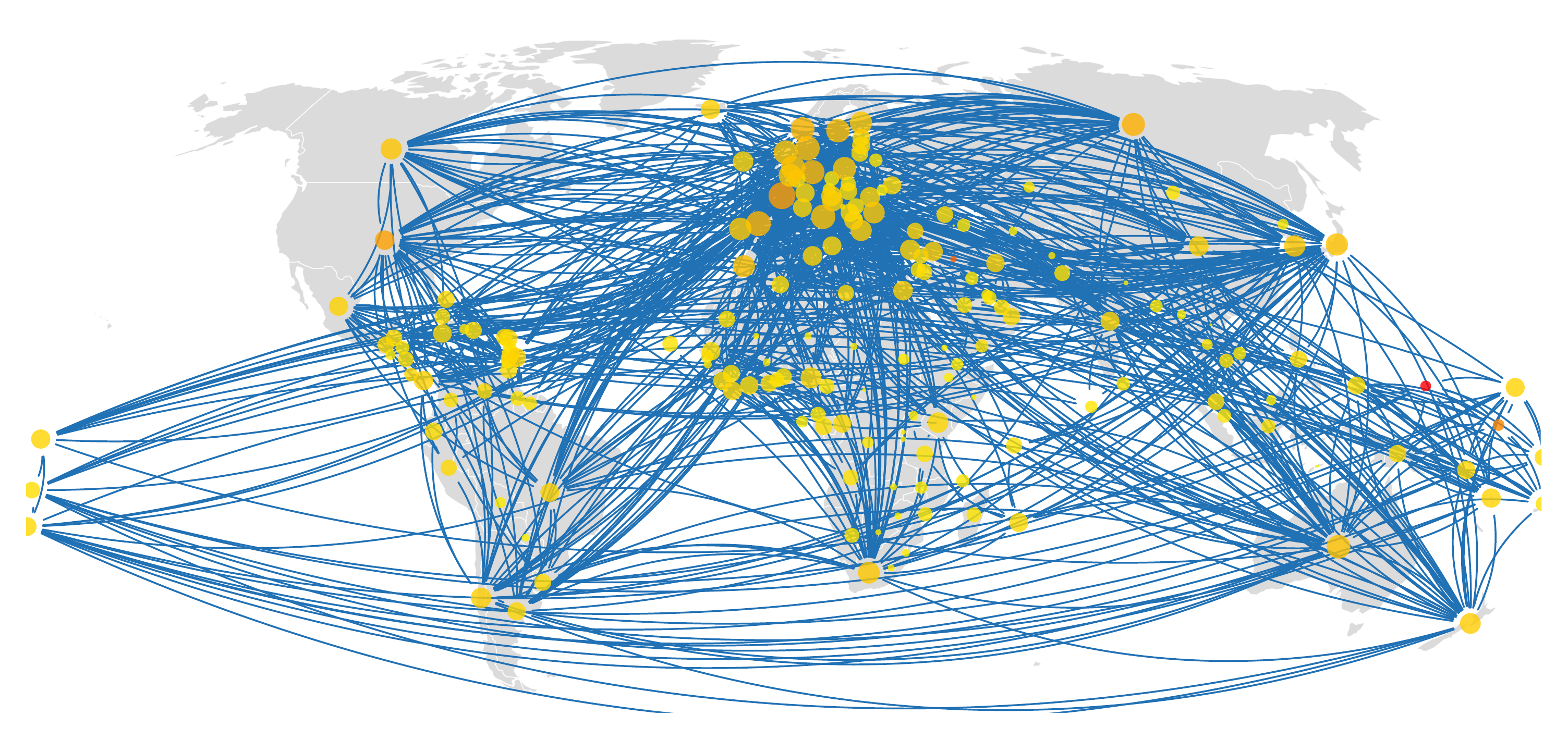}
\caption{Sea and fisheries}
\label{fig:map_sea}
\end{subfigure}
\begin{subfigure}[c]{0.46\textwidth}
\centering
\includegraphics[width=\textwidth]{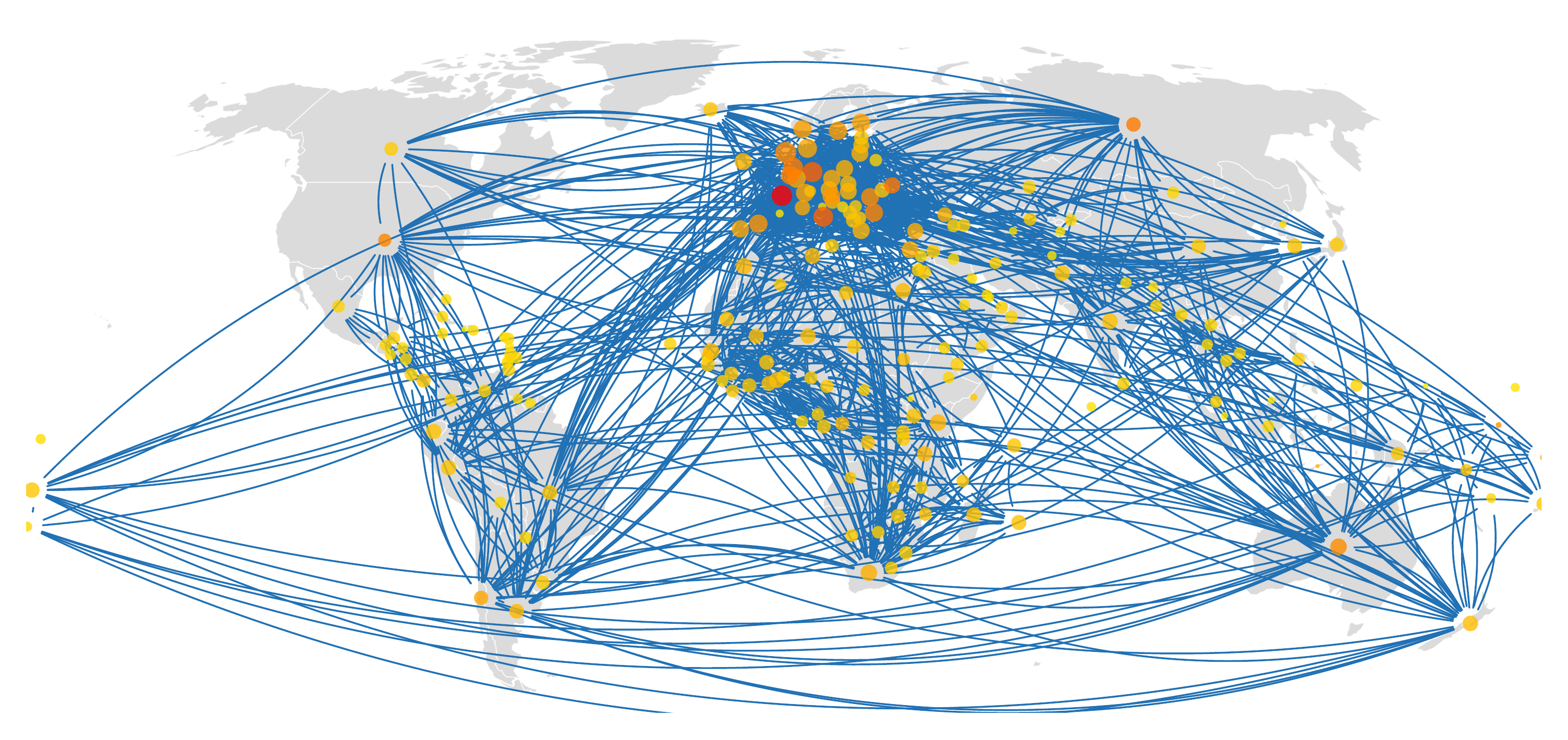}
\caption{Wild species and ecosystems}
\label{fig:map_species}
\end{subfigure} \\
\begin{subfigure}[c]{0.46\textwidth}
\centering
\includegraphics[width=\textwidth]{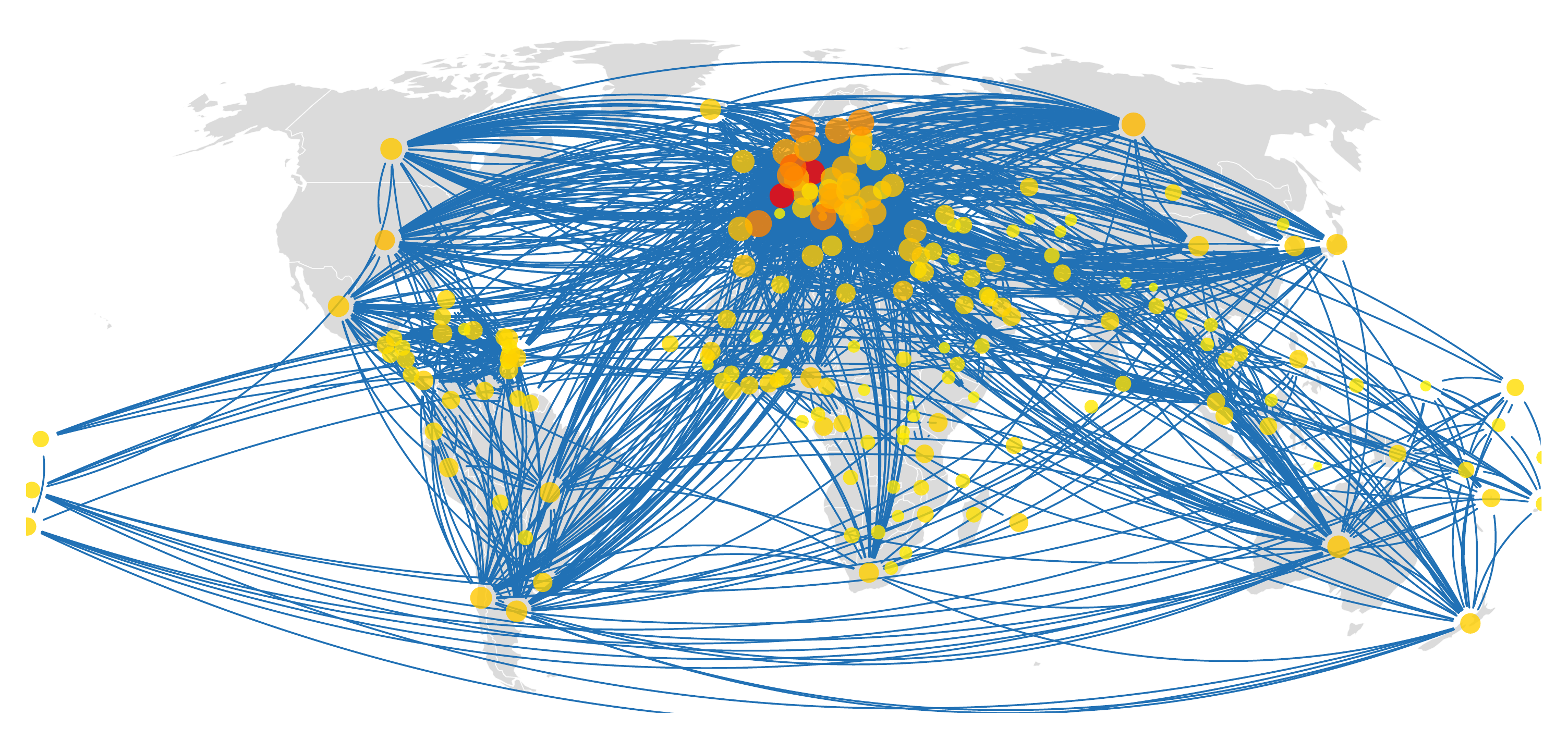}
\caption{Waste and hazardous substances}
\label{fig:map_waste}
\end{subfigure} 
\begin{subfigure}[c]{0.46\textwidth}
\centering
\includegraphics[width=\textwidth]{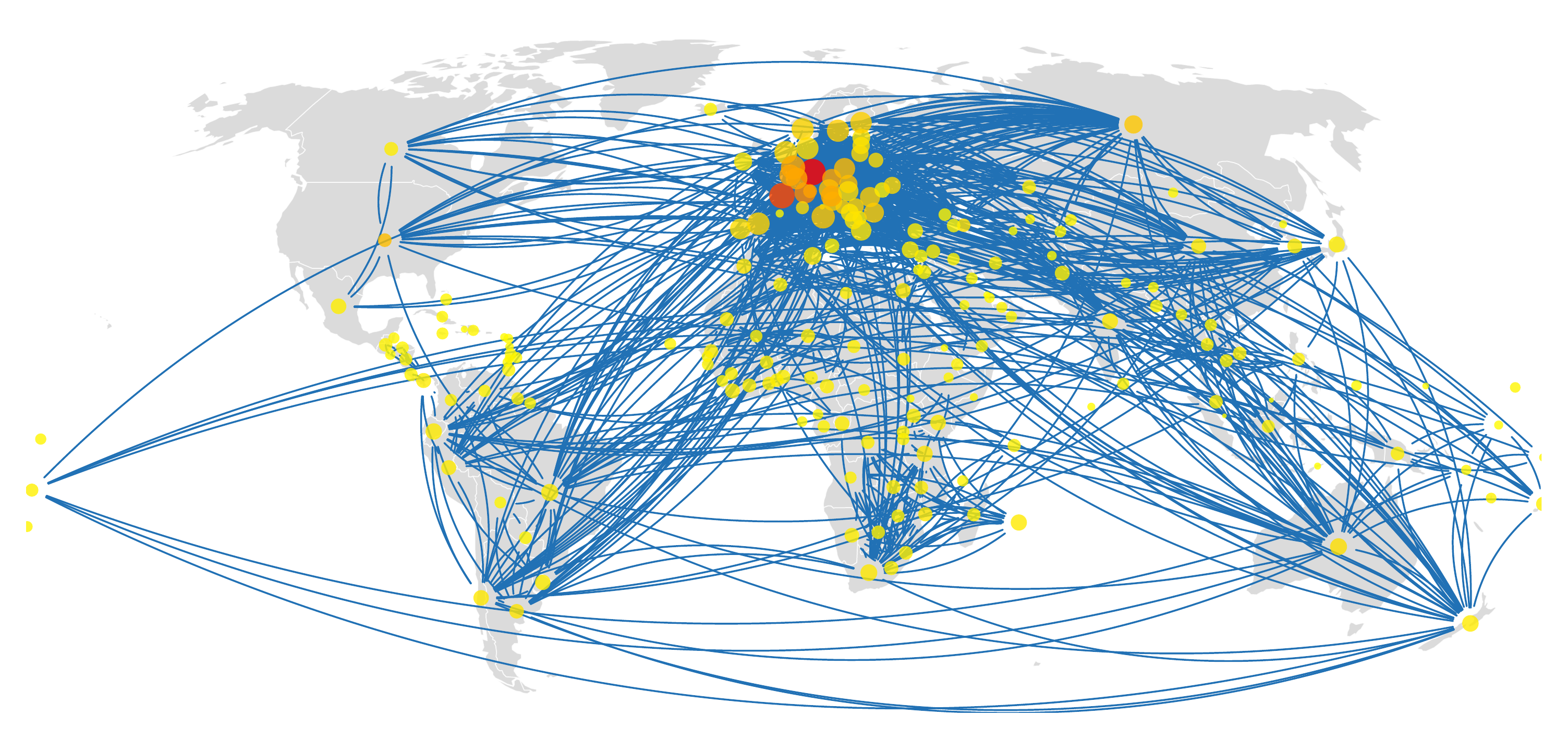}
\caption{Natural resources}
\label{fig:map_resource}
\end{subfigure} \\
\begin{subfigure}[c]{0.46\textwidth}
\centering
\includegraphics[width=\textwidth]{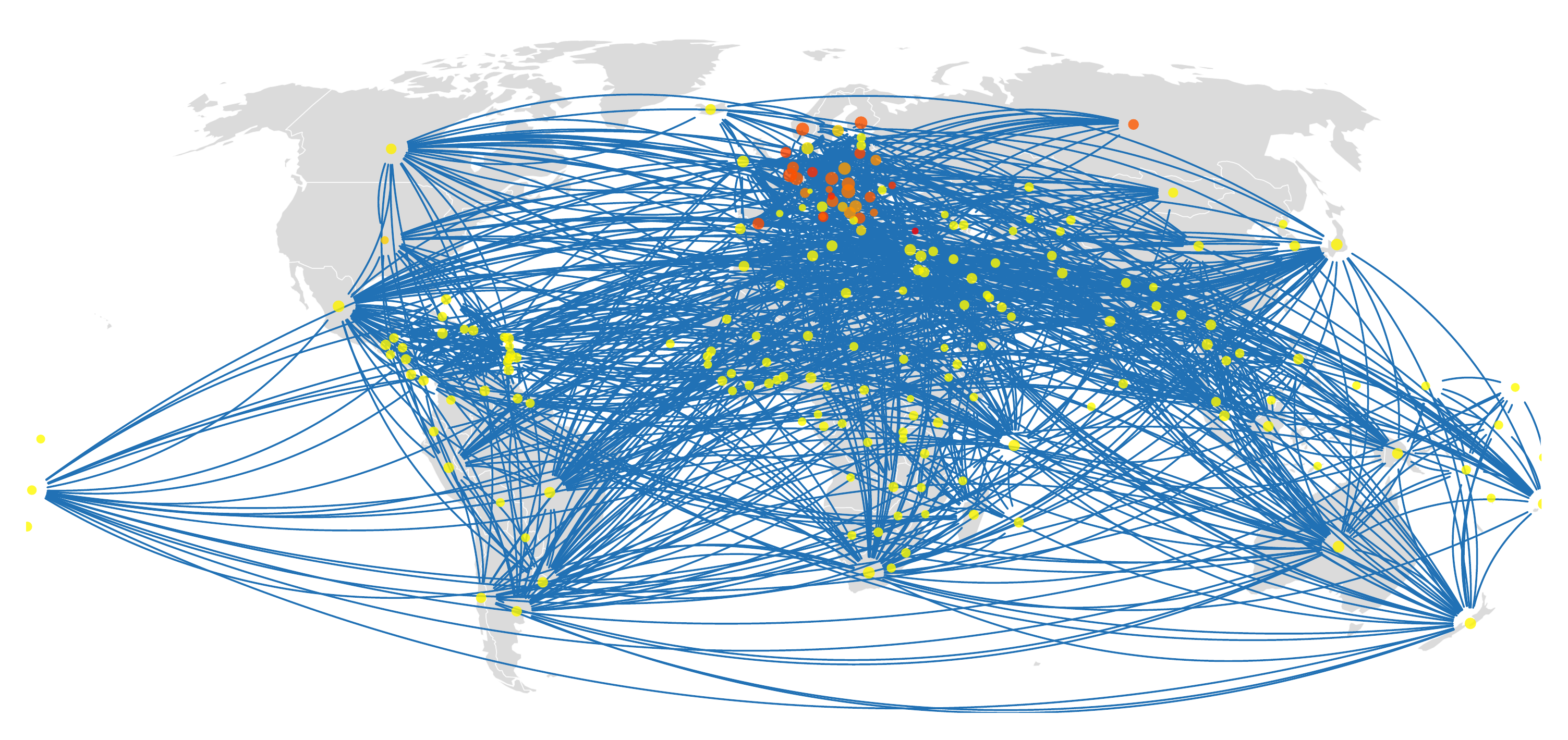}
\caption{Air and atmosphere}
\label{fig:map_air}
\end{subfigure} 
\caption{Country networks for different treaty categories in 2015.}

\begin{flushleft}\textit{Note: For the sake of visualisation, the figure only shows the top 10 percent of links in terms of weight. The size of a node is proportional to its strength, and the colour of a node (red = stronger; yellow = weaker) reflects its weighted local clustering coefficient measured using the method proposed by~\cite{onnela2005intensity}.}\end{flushleft}
\label{fig:map_subjects}
\end{figure}

Consistent with the prominence of global treaties, the cooperation relations on air and atmosphere are distributed evenly across the map and do not have an obvious core (Fig.~\ref{fig:map_subjects}). This is in contrast to most other subject areas, where Fig.~\ref{fig:map_subjects} shows a prominent core located in Europe. 

A further result of note concerns the role of the UN in air and atmosphere treaties. Unlike in the other categories, we cannot construct a statistically significant network when excluding UN-sponsored treaties. In other words, in the area of air and atmosphere there are no significant cooperation relationships among countries without the support of the UN. The results confirm that the UN has been an an effective facilitator in promoting cooperation on issues such as ozone layer depletion, climate change, and air pollution.

\subsection{Local clustering and node degree}

It is instructive to look at the inter-relationship between different network metrics. We first focus on the correlation between unweighted local clustering and degree. 

For the cooperation networks on species, waste, and natural resources, countries with a larger degree tend to have a smaller local clustering coefficient: there is a statistically significant and negative Pearson correlation coefficient between degree and local clustering coefficient. This is consistent with a hierarchical structure in which small clusters are densely connected and combine to form larger, but less dense, groups~\citep{ravasz2003hierarchical}. Similarly, when coping with these environmental issues, countries with a large number of partners are less involved in interconnected closed triplets.

In contrast, neither the cooperation network on sea and fisheries nor the network on air and atmosphere appear to have a hierarchical structure. In these networks, countries with a high local clustering coefficient also have a high degree: the Pearson correlation coefficient between the two metrics is statistically significant and positive. 

\begin{figure}[H]
\centering
\includegraphics[scale=0.3]{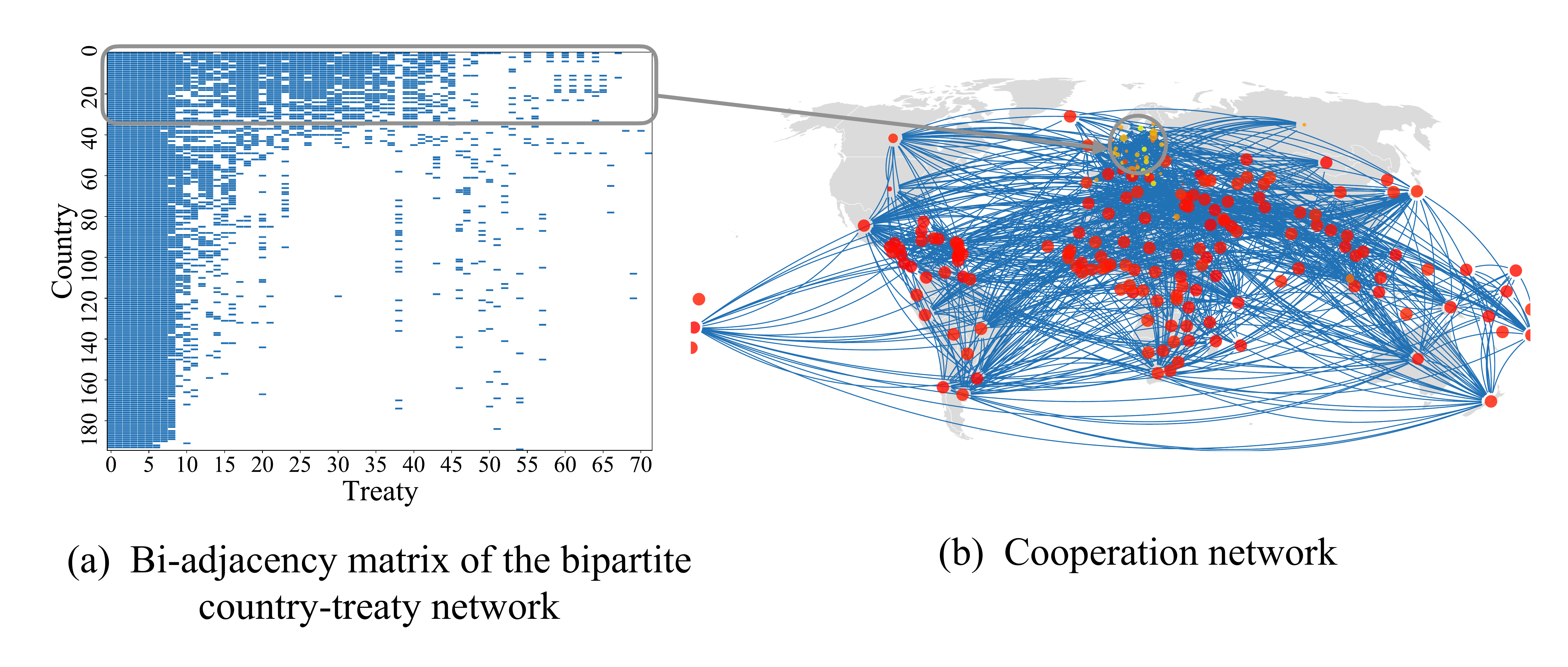}
\caption{The cooperation network and the bi-adjacency matrix of the bipartite country-treaty network for air and atmosphere in 2015.}
\begin{flushleft}\textit{Note: In panel (a), the rows (treaties) and the columns (countries) of the bi-adjacency matrix have been sorted by degree. The figure highlights the group of few countries that signed many treaties and ended up with small values of degree and clustering. In panel (b), for the sake of visualisation, the country network only shows the top $10$ percent of links in terms of weight. The size of a node is proportional to its degree, and the colour of a node reflects its unweighted local clustering coefficient measured (red = stronger; yellow = weaker).}\end{flushleft}
\label{fig:matrix_map_air}
\end{figure}

The positive coefficient observed in the air and atmosphere network seems to result of countries falling into two distinct groups with different treaty-related behaviours. The split can be visually detected from the bi-adjacency matrix of the bipartite country-treaty network (Fig.~\ref{fig:matrix_map_air}), which depicts which countries have signed which air and atmosphere treaties. The rows (countries) and the columns (treaties) of the matrix have been sorted by degree, i.e., the number of treaties each country has signed up to, and the number of signatories each treaty has attracted, respectively. 

First, there is a large number of countries (from row $30$ to row $190$ in Fig.~\ref{fig:matrix_map_air}a), which have primarily signed large global treaties (e.g., on ozone layer depletion and climate change). In fact there are some countries (e.g., Bahrain, Burundi, Palau, Uzbekistan, Turkmenistan) that have signed up only to large treaties. The number of common treaties between any two countries in this group tends to be large enough to pass the significance test and result in a statistically validated link between them. This, in turn, results in a one-mode projection in which countries tend to have a high degree (large nodes in Fig.~\ref{fig:matrix_map_air}b) as they are connected to the many co-signatories of the large treaties, and at the same time a large local clustering coefficient (red nodes in Fig.~\ref{fig:matrix_map_air}b) 
as the co-signatories they are connected too are also likely to be connected with each other ~\citep{ravasz2003hierarchical}. 

Second, there is a smaller group of countries which signed up to a much larger number of treaties (about the first $30$ rows in Fig.~\ref{fig:matrix_map_air}a). On the one hand, the number of common treaties between these countries and others in the network is relatively small compared to the total number of treaties signed by these countries. This makes it less likely for countries in this group to form statistically significant links with countries in the former group. On the other hand, although these countries signed a large number of treaties, the number of common treaties between any two countries in this group was not, on average, sufficiently large to pass the significance test and resulted in a statistically validated link (as the number of treaties signed by any two countries increases, the number of common treaties between these countries also needs to increase to pass the significance test). Therefore, countries in this second group tend to have a low degree as well as a small local clustering coefficient (small and yellow nodes in Fig.~\ref{fig:matrix_map_air}b). 

To sum up, it is the distinctiveness of the treaty-related behaviours of countries dealing with air and atmosphere issues that can (at least partly) explain the non-hierarchical organisation of these countries and the unusual positive correlation between degree and clustering observed in this network. 

\subsection{Local clustering and node strength}

We now turn to the correlation between the weighted local clustering coefficient and node strength. When clustering is computed through the method proposed by \cite{onnela2005intensity} (see details in \ref{subsection: method}), there is a statistically significant and positive correlation between weighted local clustering coefficient and strength for each treaty category. This implies that, when copying with environmental issues, countries characterised by intense collaborative links tend to be connected with other countries that also collaborate with each other. That is, countries at the centre of strong triplets are more likely to be embedded in closed structures, rich in closed triangles, than countries at the centre of weak triplets (see Fig.~\ref{fig:map_subjects}).  

For the air and atmosphere network there is a statistically significant and negative correlation between weighted local clustering coefficient and degree. Despite the fact that, when dealing with these environmental issues, countries with many collaborators tend to be included in triangles, the weights of the collaborative links in these triangles are small, once again because on these issues many countries tend to co-sign only very large treaties~\citep{newman2001scientificb}.

By contrast, when we compute the weighted local clustering coefficient using the method proposed by \cite{barrat2004architecture} (see details in \ref{subsection: method}), there is a negative correlation between weighted local clustering coefficient and node strength for wild species and ecosystems, waste and hazardous substances, natural resources and air and atmosphere. As this method depends on the local distribution of weights of links, these results suggest that countries involved in many intensive collaborations tend to close up only triplets made of links of relatively low intensity (compared to the average intensity). That is, according to the method proposed by \cite{barrat2004architecture}, weaker collaborations of highly collaborative countries are more likely to involve partners that collaborate themselves than the more intensive collaborative links. 

 \section{Conclusions} \label{Sec:Conclusions}
 
 Global environmental governance  has been the subject of intense academic scrutiny. This paper adds a novel angle to this debate by providing quantitative evidence from network analysis.
 
  Network analysis provides a systematic, quantitative analytical lens that can corroborate or refute evidence, often of a qualitative nature, from the existing literature. Network metrics can help to assess the depth of environmental cooperation an flush out interesting patterns of heterogeneity, such as differences by subject areas or the strategic importance of particular countries.
 
The aim of this paper was to demonstrate the power of network analysis by identifying salient network features, or stylised facts, of international environmental cooperation. The stylised facts speak directly to important debates in the political science, international relations and economics literature, providing quantitative evidence in support of several theories and conjectures that have been circulating in the literature. 

The stylised facts also suggest a rich agenda of follow-up research. There are intriguing topological differences, for example, between environmental subject areas, which are worthy of further investigation. Other lines of enquiry could move from the global metrics used here to the meso level, investigating for example the tendency of the most well-connected countries to generate exclusive collaborative groups. Another avenue for future research would study the role played by the network of IEAs in policy diffusion.  

Different networks may have to be constructed for different research questions. The network constructed here takes a country perspective. Country nodes are connected through treaty links. It is an obvious choice for an analysis interested in the international relations and political economy of environmental cooperation. Other research questions may require a treaty-based perspective, that is, a network with treaties as the nodes. They could be linked through shared signatories \citep{bohmelt2016interaction,kim2020global}, textual citations \citep{kim2013emergent,hollway2016multilevel}, content similarity \citep{hollway2016multilevel}, or geographic proximity \citep{hollway2016multilevel}.

Through judicious network design, network analysis can account for many of the rich historical, cultural and economic links that exist between countries and which go beyond joint treaty membership, potentially including also measures of soft power. 

As we hoped to demonstrate with this paper, this makes network analysis a powerful complement to the traditional tools used in the study of global governance and international environmental cooperation.

%\appendix 
%\section{} \label{appendix}

%The eleven components in the cooperation network in 1968:

%\noindent 1. Sierra Leone, Malawi, Jamaica, Ghana, Venezuela, Boliv. Rep. of, Madagascar, Cambodia, Iran, Islamic Republic of, Haiti, Malaysia, Senegal, Trinidad and Tobago, Liberia, Thailand, Dominican Republic, Uganda \\
%2. Kuwait, Syrian Arab Republic, Lebanon, Cyprus, Morocco, Tunisia, Turkey\\
%3. Indonesia, Korea Republic of, Myanmar \\
%4. Norway, Denmark, Finland \\
%5. Nicaragua, Brazil \\
%6. Pakistan, India \\ 
%7. Romania, Bulgaria \\
%8. Sri Lanka, Lao People's Dem. Rep.\\
%9. Ecuador, Peru\\
%10. Côte d'Ivoire, Algeria \\
%11. Australia, New Zealand \\

%There are $36$ countries that were not in the cooperation network in 1979: Namibia, Croatia, Slovakia, Palestinian Authority, Bhutan, Timor-Leste, Tajikistan, Uzbekistan, Montenegro, Kyrgyzstan, Vanuatu, Palau, Micronesia, Fed. States, Antigua and Barbuda, Georgia, Liechtenstein, Turkmenistan, South Sudan, Latvia, Macedonia FYR, Moldova Republic of, Bosnia and Herzegovina, Armenia, Estonia, Kazakhstan, Zimbabwe, Azerbaijan, Brunei Darussalam, Slovenia, Eritrea, Lithuania, Czech Republic, Andorra, Marshall Islands, Taiwan, Hong Kong.

\newcommand{\sym}[1]{{#1}} 

% \scriptsize
% unsrtnat

\newpage

\bibliography{references}

\end{document}